\documentclass[3p, authoryear, times, 12 pt]{elsarticle}

\usepackage{amsmath, amssymb, amsthm}
\usepackage[colorlinks = true]{hyperref}
\usepackage{graphicx, multirow}
\usepackage{pgfplots, pgfplotstable}
\usepgfplotslibrary{external, statistics}
\usepackage{tikz}
\usetikzlibrary{shapes, arrows, positioning, plotmarks, patterns}

\pgfplotsset{colormap/bluered}

\definecolor{turquoise}{RGB}{0, 120, 200}

\AtBeginDocument{%
\hypersetup{
linkcolor = turquoise,
citecolor = turquoise,
filecolor = turquoise,
urlcolor = turquoise,
bookmarks = true,
pdftoolbar = true,
pdfmenubar = true,
pdfstartview = {FitH},
pdftitle = {A Probabilistic Approach to Driver Assistance for Delay Reduction at Congested Highway Lane Drops},
pdfauthor = {Goodarz Mehr},
pdfsubject = {Lane Drop Advisory},
}}

\begin{document}
	
	\title{A Probabilistic Approach to Driver Assistance for Delay Reduction \\ at Congested Highway Lane Drops}
	
	\author[VT]{Goodarz Mehr\corref{cor1}}
	\ead{goodarzm@vt.edu}
	
	\author[VT]{Azim Eskandarian}
	\ead{eskandarian@vt.edu}
	
	\cortext[cor1]{Corresponding author}
	
	\address[VT]{Department of Mechanical Engineering, Virginia Tech, Blacksburg, VA 24061, USA}
	
	\begin{abstract}
		
		This paper proposes an onboard advance warning system based on a probabilistic prediction model that advises vehicles on when to change lanes for an upcoming lane drop. Using several traffic- and driver-related parameters such as the distribution of inter-vehicle headway distances, the prediction model calculates the likelihood of utilizing one or multiple lane changes to successfully reach a target position on the road. When approaching a lane drop, the onboard system projects current vehicle conditions into the future and uses the model to continuously estimate the success probability of changing lanes before reaching the lane-end, and advises the driver or autonomous vehicle to start a lane changing maneuver when that probability drops below a certain threshold. In a simulation case study, the proposed system was used on a segment of the I-81 interstate highway with two lane drops - transitioning from four lanes to two lanes - to advise vehicles on avoiding the lane drops. The results indicate that the proposed system can reduce average delay by up to 50\% and maximum delay by up to 33\%, depending on traffic flow and the ratio of vehicles equipped with the advance warning system.
		
		\copyright 2020. Manuscript accepted for publication in the International Journal of Transportation Science and Technology. This manuscript version is made available under the CC-BY-NC-ND 4.0 license (\href{http://creativecommons.org/licenses/by-nc-nd/4.0/}{http://creativecommons.org/licenses/by-nc-nd/4.0/}).
	
	\end{abstract}
	
	\maketitle
	
	\section{Introduction} \label{Section1}
	
	Lane changes are essential to highway driving, yet they can have a deteriorating effect on traffic flow and safety. These maneuvers are influenced by several factors, including driving behavior, state of nearby vehicles, and the urgency to change lanes \citep{Brackstone}. Because of this, any small mistake or unsafe driving behavior can result in an accident. In the United States, between four to ten percent of all reported motor vehicle crashes are due to unsafe lane changes. In addition to the fatalities, traffic delay caused by these accidents incurs an economic loss \citep{Sen, Lisheng, Dijck}. Such incidents and associated delays can be partially mitigated by providing drivers with timely information of the road ahead and using assistant systems that help control the vehicle. \par
	
	Lane changes can be classified as either discretionary or mandatory \citep{Zhang1, Vechione}. Discretionary lane changes are generally performed to overtake slow traffic and move to a faster lane. In contrast, drivers perform mandatory lane changes to follow a planned route or avoid an obstacle, for example a lane drop. Compared to discretionary lane changes, the impact of mandatory lane changes on traffic can be disruptive. They can deteriorate traffic safety \citep{Ahammed, Li} and cause traffic oscillation \citep{Sarvi}, traffic breakdown \citep{Lv}, and capacity drops \citep{Cassidy1}. Mandatory lane changes caused by lane drops have been shown to affect traffic flow in all lanes upstream by generating perturbation density waves \citep{Munjal}, forming queues \citep{Bertini}, and stop-and-go traffic patterns \citep{Zhang2, Yuan}. \par
	
	To manage upstream traffic at a lane drop, past studies have focused on variable speed limit (VSL) strategies \citep{Jin, Yu2, Yu1}, congestion assistants \citep{VanDriel, Roncoli, Zhang3}, or a combination of both \citep{Zhang4}. In approaches involving VSL, traffic flow is managed by dynamically modifying the speed limit upstream of the bottleneck using different control methods and optimization strategies - ranging from PI- and I- controllers \citep{Jin} to genetic algorithm (GA) \citep{Yu2} and tabu search algorithm \citep{Yu1} - to mitigate a drop in capacity. While macroscopic simulations have shown that VSL strategies can achieve that goal and reduce total travel time (TTT) by up to 16\% \citep{Yu2}, as \citet{Zhang4} notes because capacity drops and delays often happen due to lane changes close to the bottleneck, absent a lane assignment strategy the system often breaks down in microscopic simulations and fails to mitigate capacity drop. To that end, centralized lane assignment systems using connected vehicle technology, either standalone \citep{Roncoli, Zhang3} or combined with VSL strategies \citep{Zhang4}, have been proposed and shown to reduce TTT by up to 40\% \citep{Zhang3}. \par
	
	In this paper we take a different approach to delay reduction at lane drops and propose an advance warning system based on a probabilistic prediction model that advises vehicles on when to change lanes for an upcoming lane drop. Compared to past studies, our approach can be implemented as a simple, individualized, real-time onboard system using readily available information, removing the need for the hardware necessary for connected vehicle technology or a centralized lane assignment system. Furthermore, as our proposed system only advises vehicles on their lane changing behavior, it can be coupled with VSL strategies to better control both the longitudinal and lateral behavior of vehicles to ensure a more orderly transition of vehicles from the blocked lane into other lanes. Finally, we introduce a microscopic simulation setup to evaluate the performance of the proposed system and test its effectiveness for different traffic flow and penetration rate (ratio of vehicles with the onboard system) conditions.
	
	The remainder of this paper is structured as follows. \autoref{Section2} presents the methodology, including a brief overview of the probability model underlying the system and the simulation setup used to evaluate its performance. \autoref{Section3} presents our findings and a discussion of the effects of the system on traffic flow and traffic efficiency. Finally, \autoref{Section4} concludes the findings of this paper.
	
	\section{Methodology} \label{Section2}
	
	A road bottleneck is a location where downstream flow capacity is lower than upstream capacity \citep{Roncoli}. This can be the result of road features like lane drops and merges, or temporary blockages such as work zones and traffic accidents. \par
	
	At a lane drop bottleneck, the nominal capacity is the maximum traffic flow that can be maintained downstream if the upstream traffic flow is no larger than that capacity. In other words, if upstream capacity for a road transitioning from $n$ lanes to $n - 1$ lanes, $n \ge 2$, is denoted by $C$, then the bottleneck capacity is $C_{b} = \frac{n - 1}{n}C$. However, if upstream traffic flow is larger than $C_{b}$ or if lane changing maneuvers of vehicles in the blocked lane trying to get in other lanes cause vehicles to decelerate and disrupt traffic, the actual capacity, denoted by $C_{a} = \delta C$, is lower than the nominal capacity $C_{b}$. This reduction in capacity is called capacity drop and past studies have shown that during congestion actual capacity can be anywhere from 5\% to 20\% lower than the nominal capacity \citep{Cassidy2, Chung}. Capacity drop causes a disruption in traffic and results in higher delays for all vehicles. \par
	
	To prevent or postpone capacity drop and reduce delay at lane drop bottlenecks, we propose an advance warning system based on a model that predicts the likelihood of reaching a near-term target position using one or multiple lane changes \citep{Mehr1} which has previously been shown to be effective at reducing traffic delay at highway diverges \citep{Mehr2}. Using this model and driving conditions at each moment, vehicles in the blocked lane equipped with the system constantly calculate the probability of leaving that lane under those driving conditions before reaching the lane end. They change lanes when that probability drops below a certain threshold. \par
	
	The proposed system is implemented in VISSIM\textsuperscript{\textregistered} for a simulation of a section of the I-81 interstate highway transitioning from four lanes to two lanes to evaluate its performance in reducing delay under different traffic flows and penetration rates. In what follows, \autoref{Section2.1} gives a brief overview of the probability model introduced above and \autoref{Section2.2} describes our implementation of the proposed advance warning system and the simulation setup used to evaluate its effectiveness.
	
	\subsection{Probability model} \label{Section2.1}
	
	\citet{Mehr1} introduced a model to estimate the probability that a vehicle can successfully perform one or multiple lane changes to reach a near-term goal state. While a brief overview of the model is provided here for the sake of completeness, the reader can find the complete model derivation and validation at \citet{Mehr1}.
	
\begin{figure}[ht!]
		\centering
		\includegraphics[width = \columnwidth]{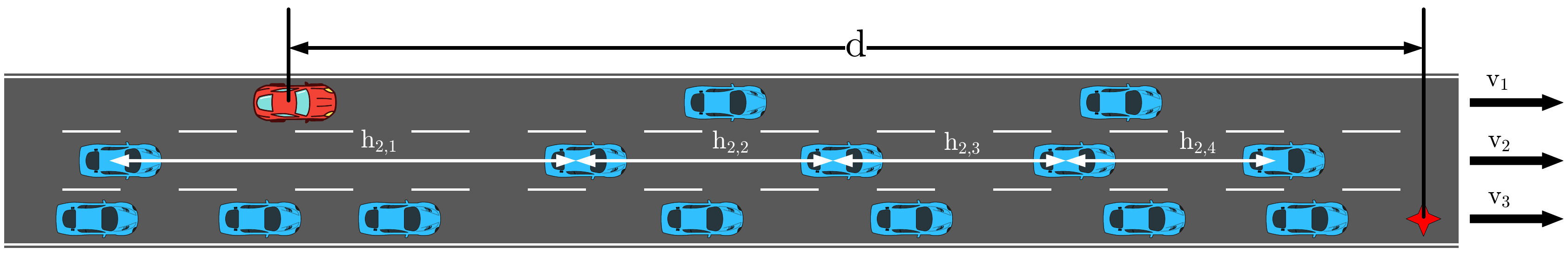}
		\caption{Notations used in this paper for a road segment with three lanes. The red car is the ego vehicle and the red star shows the goal state. The model estimates the probability that the red car can reach the red star if it does not change any driving conditions (velocity, lane changing behavior, etc.)} \label{Figure1}
	\end{figure}

	We assume, without loss of generality, that the road has $n$ lanes numbered from left to right by 1 to $n$. We further assume that from its current position on lane 1, the ego vehicle intends to reach a position that is a distance $d$ ahead on lane $n$. Let $P(S)$ denote the probability of successfully doing so. The model proposed in \citet{Mehr1} estimates $P(S)$ by making a few assumptions. First, it assumes that on each lane $i$, $1 \le i \le n$, the velocity of all vehicles is equal to $v_{i}$, obtained from averaging the velocity of all vehicles on that lane over a certain time period. Second, the model assumes that on each lane $i$, inter-vehicle headway distances (rear bumper to rear bumper) are independent identically distributed (i.i.d.) random variables sharing a common log-normal distribution with parameters $\mu_{i}$ and $\sigma_{i}$ \citep{Mei}. Finally, the model assumes that a Gipps gap acceptance model \citep{Gipps} can approximate ego vehicle lane changes. That is, the ego vehicle moves from lane $i - 1$ to lane $i$ only if the gap between its leading and trailing vehicles on lane $i$ is at least as large as a critical gap $g_{i}$. It takes $t_{i}$ seconds to complete such a lane change. These assumptions are illustrated in \autoref{Figure1}.
	
	For the case outlined above we can denote $P(S) = f_{n}(d, v_{1 : n}, \mu_{2 : n}, \sigma_{2 : n}, g_{2 : n}, t_{2 : n})$, where $a_{r : s}$ means $a_{r}, a_{r + 1}, \ldots, a_{s}$ for any parameter $a$ and indices $r \ge s$. Estimation of $P(S)$ is done recursively, starting from the base case of $n = 2$. For $n = 2$, $P(S)$ is calculated through interpolation of values from a look-up table, because the probability does not have a closed-form expression. The look-up table was obtained through Monte Carlo simulations of the problem normalized for unit distance. For $n > 2$, $P(S)$ is recursively calculated by
	\begin{equation} \label{Equation1}
		\begin{split}
			f_{n}(d, v_{1 : n}, &\mu_{2 : n}, \sigma_{2 : n}, g_{2 : n}, t_{2 : n})\\
			&= \int_{0} ^ {d} f_{2}(d - x, v_{n - 1 : n}, \mu_{n}, \sigma_{n}, g_{n}, t_{n})\frac{\partial}{\partial x} f_{n - 1}(x, v_{1 : n - 1}, \mu_{2 : n - 1}, \sigma_{2 : n - 1}, g_{2 : n - 1}, t_{2 : n - 1})\mathrm{d}x\\
			&= \frac{\partial}{\partial x}\int_{0} ^ {d} f_{2}(d - x, v_{n - 1 : n}, \mu_{n}, \sigma_{n}, g_{n}, t_{n})f_{n - 1}(x, v_{1 : n - 1}, \mu_{2 : n - 1}, \sigma_{2 : n - 1}, g_{2 : n - 1}, t_{2 : n - 1})\mathrm{d}x,
		\end{split}
	\end{equation}
	which is obtained from the law of total probability \citep{Leon}. Extensive VISSIM\textsuperscript{\textregistered} traffic simulations for a range of multiple parameters showed that in most cases the model is accurate to within 4\% of the actual probability \citep{Mehr1}.
	
	\subsection{Simulation setup} \label{Section2.2}

	The proposed advance warning system uses the probability model to advise vehicles on when to change lanes to reach a particular goal state, here avoiding a lane drop. Specifically, when a vehicle approaches a lane drop and is in the blocked lane, the system uses traffic data and vehicle conditions (speed and distance to the lane-end) to constantly calculate the probability of reaching the adjacent lane before the lane-end point under those conditions and advises the vehicle to change lanes when that probability drops below a certain limit. If a large portion of vehicles use this system, it can help them change lanes at the proper moment to reduce overall traffic delay. \par
	
	VISSIM\textsuperscript{\textregistered} traffic simulations were used to evaluate the performance of the proposed system in reducing delay at a highway segment with two consecutive lane drops. Simulations were performed for a variety of traffic conditions, obtained by changing peak traffic flow and proposed system penetration rate. For each case, we studied how different thresholds for the probability model (the value at which it advises the driver to start changing lanes) affect traffic behavior and average delay. Details of the simulation setup are presented in \autoref{Section2.2.1} to \autoref{Section2.2.4}.
	 
	\subsubsection{Simulation fundamentals} \label{Section2.2.1}
	
\begin{figure}[ht!]
		\centering
		\includegraphics[width = 0.48\columnwidth]{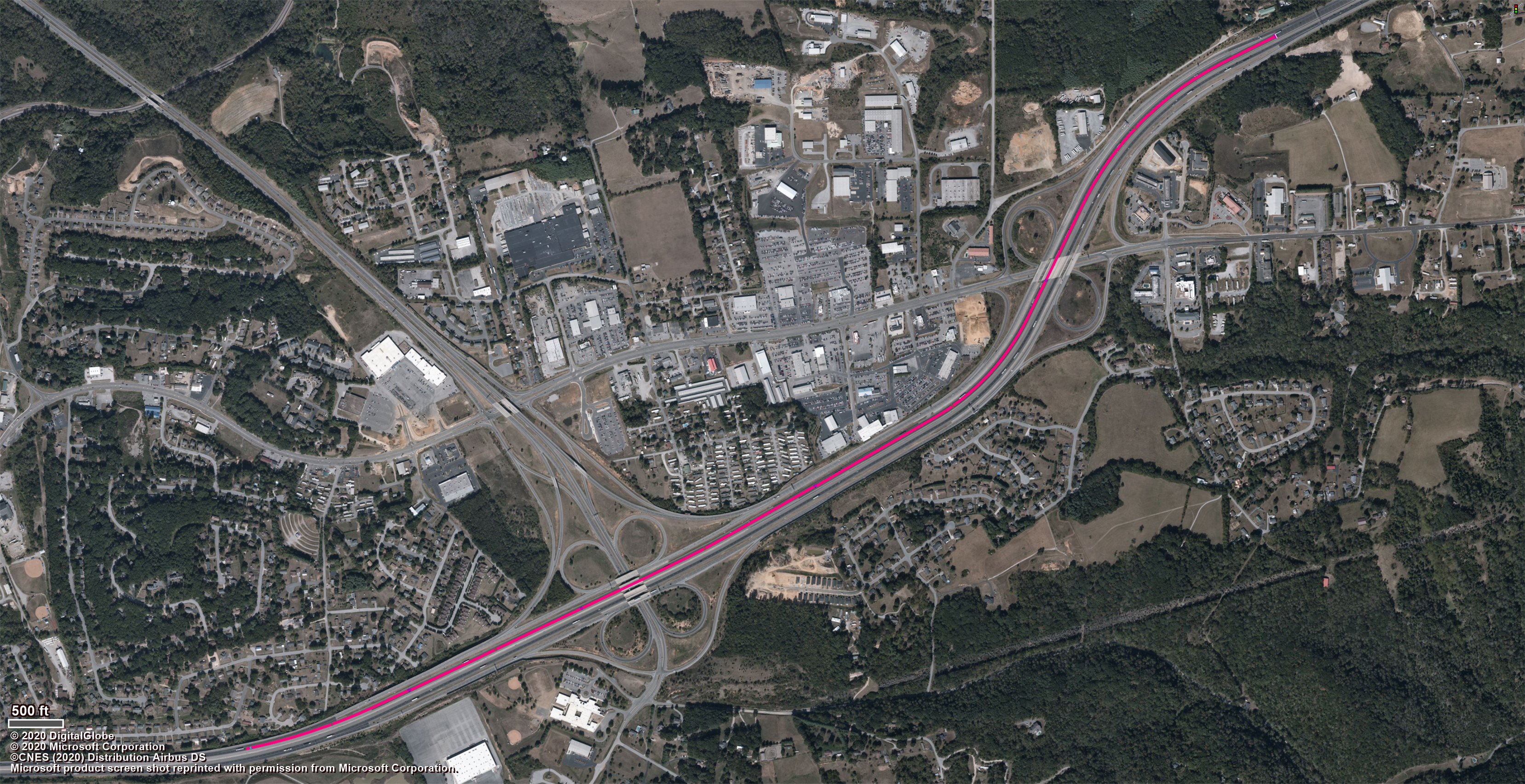}
		\includegraphics[width = 0.48\columnwidth]{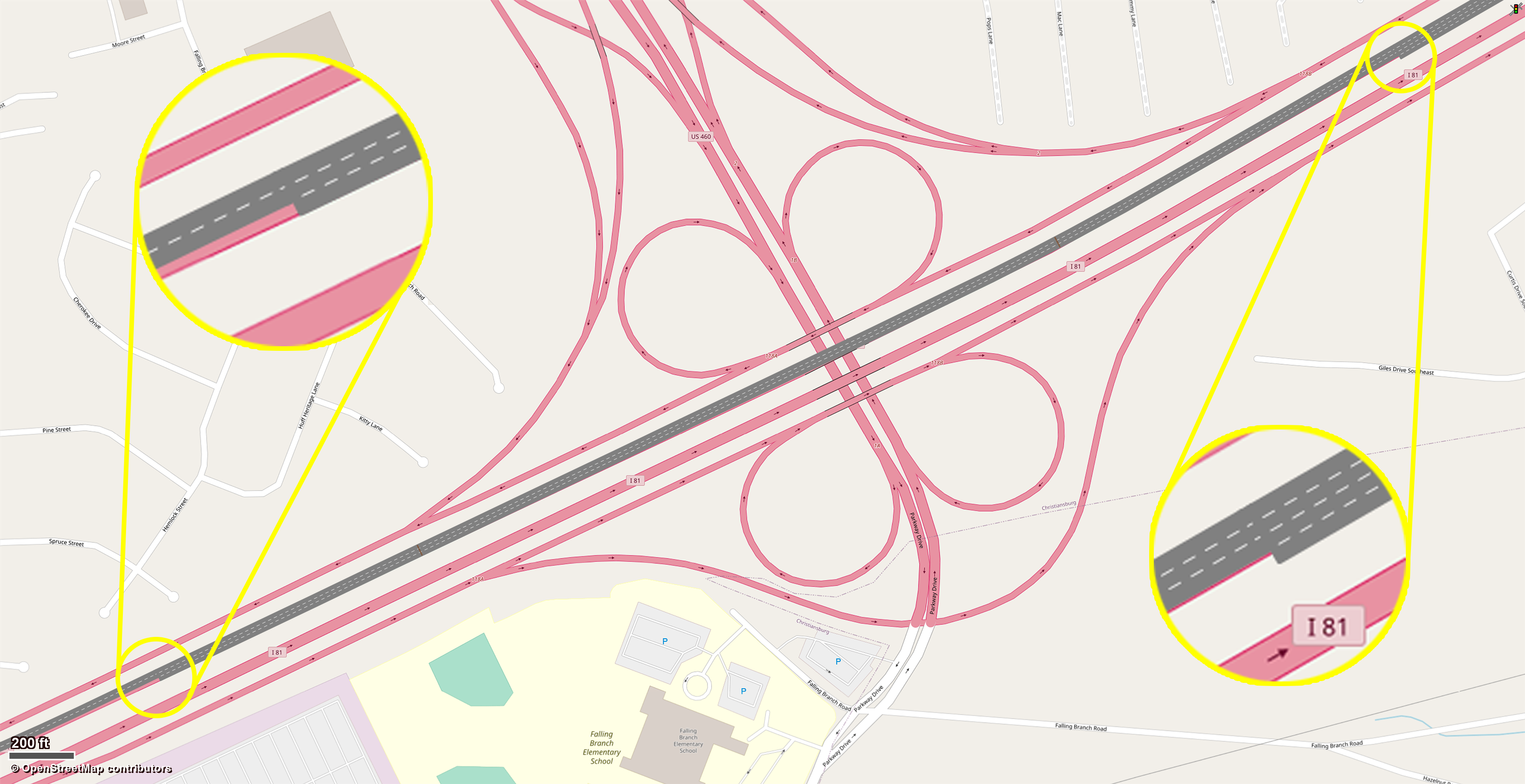}
		\caption{Left: Bird's-eye view of the southbound I-81 highway segment used for traffic simulations. The segment starts on the top right corner of the image and ends on the bottom left corner. It is 12,210.91 ft (2.31 mi) long and transitions from four lanes to two lanes via two consecutive lane drops. \\ Right: A section of the simulated highway segment. It shows the highway transitioning from four lanes to two lanes via two consecutive lane drops.} \label{Figure2}
	\end{figure}

	Traffic simulations were performed on a segment of the southbound I-81 interstate highway near Blacksburg, Virginia, shown in the left image of \autoref{Figure2}. The segment is 12,210.91 ft (2.31 mi) long and transitions from four lanes to two lanes via two consecutive lane drops, as shown in the right image of \autoref{Figure2}. It starts just after Exit 118 and ends just before the merge from 118A. It has one vehicle input and one vehicle output, located at either ends of the segment. The posted speed limit along the segment is 70 mph (roughly 112.7 km/h), though actual speeds vary based on traffic. \par
	
	The simulated road segment was divided into five sections (links) for better assignment of driving behavior, with the endpoint of each link the same as the start of the next link. The first link started from the beginning of the segment and ended slightly before the first posted lane drop sign for the first lane drop, while the second link ended at the middle of the first lane drop taper. Similarly, the third link ended slightly before the first posted lane drop sign for the second lane drop, while the fourth link ended at the middle of the second lane drop taper. The last link ended just before the merge from 118A. Link details are shown in \autoref{Table1}.
	
	\begin{table}[ht!]
		\renewcommand{\arraystretch}{1.1}
		\caption{Length and number of lanes of links used for simulation.} \label{Table1}
		\centering
		\resizebox{0.44\columnwidth}{!}{
		\begin{tabular}{c c c}
			\hline
			Link number & Length (ft) & Number of lanes \\
			\hline
			1 & 3275.312 & 4 \\
			2 & 2998.360 & 4 \\
			3 & 2490.507 & 3 \\
			4 & 1798.360 & 3 \\
			5 & 1632.506 & 2 \\
			\hline
		\end{tabular}}
	\end{table}
	
	Simulations were conducted according to Virginia Department of Transportation's (VDOT) Traffic Operations and Safety Analysis Manual (TOSAM) and VISSIM\textsuperscript{\textregistered} User Guide \citep{Tosam, VissimGuide}. The latter recommends running each simulation case 10 times with different - but consistent - random seeds and then averaging the results, but given our observation that for each case a few runs would crash before finishing\footnote{During some runs, when the advance warning system sent a signal for a vehicle to change lanes, VISSIM\textsuperscript{\textregistered}'s internal model that controls driving behavior calculated a trajectory angle larger than 90 degrees, resulting in a crash.}, each case was run 16 times to satisfy this recommendation. For each case the runs started from a random seed of 42 with an increment of 5 for each following run. Each run was set for 9000 simulation seconds. The first 1800 seconds were the seeding period and the following 7200 seconds were the analysis period, with the peak period defined as the time between 3600 and 7200 seconds. Simulation parameters were set for each 900 second interval \citep{VissimGuide}. \par
	
	For all cases, input vehicle flow $q_{i}$ was set to 2400 vehicles per hour (veh/hr) for the entirety of the simulation (obtained from historical flow data) except for the peak period where $q_{i, p}$ was one of the parameters studied for its effects on overall system performance. $q_{i, p}$ was set to either 4400, 4600, or 4800 veh/hr, modeling hypothetical temporary, rush-hour traffic (for example AM or PM peak flow or traffic after a football game at the nearby Lane Stadium) \citep{I81, AADT}. Traffic consisted of entirely North American vehicles \citep{VissimGuide} (slightly larger than European vehicles commonly used by default in VISSIM\textsuperscript{\textregistered}, matching the composition of vehicles on the road in North America) divided into three different vehicle types: cars, smart cars, and heavy goods vehicles (HGVs). Cars and Smart Cars were identical, with the exception that an external driver model (EDM) controlled the lane change initiation behavior of Smart Cars, modeling vehicles that use the proposed system (see \autoref{Section2.2.3}). \par
	
	Vehicle composition varied from case to case and during the peak period. The ratio of HGVs was set to 15\% during the peak period and to 20\% at all other times \citep{I81, AADT}. The ratio of smart cars $r$ (i.e. the penetration rate of the proposed system) was another parameter that was studied. For each value of $q_{i, p}$, $r$ was set to either 10\%, 40\%, or 70\%, resulting in 9 overall cases. The rest of the vehicles were simply cars. As an example, for a case with $r = 40\%$, the ratio of cars, smart cars, and HGVs was set to 45\%, 40\%, and 15\% during the peak period and to 40\%, 40\%, and 20\% at all other times, respectively. All vehicle types had a desired speed distribution of 70 mph at the input, which meant assigning a desired speed between 67 mph and 80 mph to each vehicle at random with uniform probability \citep{Tosam}. \par
	
	To record traffic data during the simulation, sets of data collection points were defined at the midpoint of each link. They recorded time-stamped velocity of vehicles passing through them, modeling real-world loop detectors. In addition, defining a set of travel time measurements allowed us to measure delays and total travel times (TTT) during the simulation. The measurement started at the beginning of the road segment and ended near its end, covering a total distance of 12,178.38 ft.
	
	\subsubsection{Driving behavior} \label{Section2.2.2}
	
	Driving behavior was defined according to \citet{Tosam, VissimGuide} using data from a previous VDOT study \citep{Fairfax} with parameter values shown in \autoref{Table2}. Per the recommendations of \citet{VissimGuide}, the default Freeway (free lane selection) driving behavior - using the Wiedemann 99 driving model with default parameter values - was used for links 1, 3, and 5 where a large number of lane changes were not expected \citep{Wiedemann}. The Weave \& Merge driving behavior was used for links 2 and 4 where we anticipated significant weaving and merging, given that their starting point marked the first posted sign for each lane drop.
	
	\begin{table}[ht!]
		\renewcommand{\arraystretch}{1.1}
		\caption{Driving behavior parameters used for freeway simulation.} \label{Table2}
		\centering
		\resizebox{0.96\columnwidth}{!}{
		\begin{tabular}{c c c}
			\hline 
			Parameter & Freeway (free lane selection) & Weave \& Merge \\
			\hline
			CC0 (Standstill Distance) (ft) & \phantom{-}\phantom{0}4.92 & \phantom{-}\phantom{0}4.92 \\
			CC1 (Headway Time) (s) & \phantom{-}\phantom{0}0.9\phantom{0} & \phantom{-}\phantom{0}0.9\phantom{0} \\
			CC2 (Following Variation) (ft) & \phantom{-}13.12 & \phantom{-}13.12 \\
			\hline
			Maximum Deceleration (Own Vehicle) (ft/$\mathrm{s} ^ {2}$) & -13.12 & -15.00 \\
			Maximum Deceleration (Trailing Vehicle) (ft/$\mathrm{s} ^ {2}$) & -\phantom{0}9.84 & -12.00 \\
			Accepted Deceleration (Own Vehicle) (ft/$\mathrm{s} ^ {2}$) & -\phantom{0}3.28 & -\phantom{0}4.00 \\
			Accepted Deceleration (Trailing Vehicle) (ft/$\mathrm{s} ^ {2}$) & -\phantom{0}1.64 & -\phantom{0}3.28 \\
			Safety Distance Reduction Factor & \phantom{-}\phantom{0}0.60 & \phantom{-}\phantom{0}0.25 \\
			Maximum Deceleration for Cooperative Braking (ft/$\mathrm{s} ^ {2}$) & -\phantom{0}9.84 & -23.00 \\
			\hline
			Advanced Merging & On & On \\
			Cooperative Lane Change & Off & On \\
			\hline
		\end{tabular}}
	\end{table}
	
	An important parameter throughout the simulations was Lane Change Distance. Vehicles that anticipate changing lanes before a certain connector (e.g. to reach a diverge point), start to act on that intention when their distance to the said connector is smaller than Lane Change Distance. For lane drop connectors (the connector between links 2 and 3 and the connector between links 4 and 5) this value was increased from the default value of 656.2 ft to the distance of that connector from the first posted lane drop sign. For the first connector, that distance was 2880 ft, while for the second connector it was 1740 ft. While the default value usually satisfies urban traffic simulation requirements, it has to be increased for highway modeling because otherwise, artificial queues appear near the lane drop \citep{VissimGuide, Gomez}. Without experimental trajectory data to calibrate the model, the values used showed a good balance between preventing artificial queues at each lane drop and forcing all merging vehicles out of the blocked lane much earlier than they are supposed too.
	
	\subsubsection{External driver model} \label{Section2.2.3}
	
	Different driving behavior aspects of one group of or all vehicles in a VISSIM\textsuperscript{\textregistered} traffic simulation can be controlled using its external driver model (EDM) API. The EDM was used in this study to simulate the proposed onboard advance warning system for an upcoming lane drop. \par
	
	To create a baseline for later comparison, each case was first simulated with all vehicles using VISSIM\textsuperscript{\textregistered}'s internal model. Average values of $v_{i}, \mu_{i}$, and $\sigma_{i}$, $2 \le i \le 3$ as appropriate (based on ego vehicle's lane number), were calculated for different road segments in 900-second intervals using information from data collection points defined earlier. For example, average $v_{i}, \mu_{i}$, and $\sigma_{i}$ values for each 900-second interval for the first link were calculated using information from the first data collection point at the midpoint of the first link. In the real world, this information can be collected once, either experimentally using loop detectors or through traffic simulations, and stored locally or in the cloud for future use by onboard warning systems. In addition, we set $g_{i} = \delta v_{i} + s_{0}$ where $\delta$ = 1.6 s and $s_{0}$ = 1 m, respectively. Even though in reality several factors - including driver aggressiveness and relative position and speed of leading and trailing vehicles in the adjacent lane - affect the critical gap, making it stochastic in nature, our choice simplified the implementation of the model. In addition, by generally being larger than the actual critical gap, the chosen gap ensures unsafe lane changes do not occur \citep{Toledo}. Finally, we set $t_{i} = 3$ s, given that it is the time it takes for VISSIM\textsuperscript{\textregistered}'s internal model to complete a lane change from when it is initiated \citep{PTV}. In the real world, both $g_{i}$ and $t_{i}$ can be tuned to match the lane changing behavior of individual drivers. \par
	
	In subsequent simulations, smart cars on the two leftmost lanes of links 1 and 2 or on the leftmost lane of links 3 and 4 used the EDM for advice on when to change lanes. Along with the values of $v_{i}, \mu_{i}, \sigma_{i}, g_{i}$, and $t_{i}$, the EDM used each vehicle's velocity as $v_{1}$ and its distance to the lane-end of that lane as $d$. For example, if a vehicle was on the leftmost lane of link 1, $d$ would be set to the distance of that vehicle to the first lane drop, whereas if the vehicle was on the second leftmost lane, $d$ would be set to the distance of that vehicle to the second lane drop. The EDM was inactive when a vehicle was on the two rightmost lanes. This general approach was slightly modified by setting $v_{i + 1} = v_{i} + v_{l}$ in cases where $v_{i + 1}$ was within the range of $v_{i} \pm v_{l}$, where $v_{l}$ = 4 m/s. This was done to more accurately represent acceleration or deceleration behavior of drivers intending to initiate a lane change and looking for a suitable gap in an adjacent lane, given that \citet{Mehr1} showed that when $v_{i}$ and $v_{i + 1}$, $1 \le i \le n - 1$, are close to each other, the estimated probability experiences a large and unrealistic drop due to a large reduction in the relative traveled distance.
	
	Throughout each simulation, the EDM constantly calculated the probability of successfully avoiding the lane drop under momentary conditions for each smart car not on the two rightmost lanes. It advised vehicles to initiate a lane change if that probability dropped below a certain threshold. For each of the 9 cases, different thresholds $p_{l}$ ranging from 0.999 to 0.75 were tested to understand their effect on travel times and the flow of traffic. \par
	
	A problem we encountered when using the EDM was how VISSIM\textsuperscript{\textregistered} handled lane changes. Whenever the EDM commanded a vehicle to change lanes, that vehicle would immediately do so without checking to see if it was safe first, leading to bizarre situations where vehicles passed through each other. To tackle this problem, we built an additional mechanism within the EDM to check for safety before commanding vehicles to change lanes. The mechanism calculated the leading and trailing critical gaps given in \autoref{Equation2} and \autoref{Equation3} using the velocity of the ego vehicle relative to its leading and trailing vehicles in the adjacent lane. It then compared those gaps with actual relative distances between the ego vehicle and said vehicles \citep{Toledo}. If the critical gap was smaller than both distances, the EDM would proceed with the lane change.
	\begin{align*}
		g_{i} ^ {\mathrm{lead, cr}} &= \exp\big(1.353 - 2.700\max[0, \Delta v_{i} ^ {\mathrm{lead}}] \nonumber - 0.231\min[0, \Delta v_{i} ^ {\mathrm{lead}}] + \epsilon ^ {\mathrm{lead}}\big), \tag{2} \label{Equation2} \\
		g_{i} ^ {\mathrm{lag, cr}} &= \exp\big(1.429 + 0.471\max[0, \Delta v_{i} ^ {\mathrm{lag}}] + \epsilon ^ {\mathrm{lag}}\big), \tag{3} \label{Equation3}
	\end{align*}
	where $\epsilon ^ {\mathrm{lead}}\sim N(0, 1.112 ^ {2})$ and $\epsilon ^ {\mathrm{lag}}\sim N(0, 0.742 ^ {2})$. In the equations above, $g ^ {\mathrm{lead}}$ denotes the gap between the rear bumper of the leading vehicle in the adjacent lane and the ego vehicle's front bumper and $g ^ {\mathrm{lag}}$ denotes the gap between the front bumper of the trailing vehicle in the adjacent lane and the ego vehicle's rear bumper. Similarly, $\Delta v ^ {\mathrm{lead}}$ and $\Delta v ^ {\mathrm{lag}}$ denote the velocity of the leading and trailing vehicles in the adjacent lane relative to the ego vehicle's velocity, respectively. Finally, $\epsilon$ is a random term associated with lane utility \citep{Toledo}.
	
	\subsubsection{Data processing and evaluation} \label{Section2.2.4}
	
	The measure of effectiveness (MoE) of our choice was average delay, defined as the difference between travel time under free flow speed and actual travel time \citep{VissimGuide}. For each simulation, VISSIM\textsuperscript{\textregistered} automatically calculated individual vehicle delays using the travel time measurement defined in \autoref{Section2.2.1}. The average $m_{i}$, standard deviation $s_{i}$, and maximum delay $a_{i}$ for each run $i$, $1 \le i \le 16$, were calculated from that data. In this paper we report the average of those values over all runs for each combination of $q_{i, p}, r$, and $p_{l}$ both for the entire analysis period and for each 900-second interval. In other words, for each combination of $q_{i, p}, r$, and $p_{l}$ we report $m = \frac{1}{16}\sum_{i = 1} ^ {16} m_{i}, s = \frac{1}{16}\sum_{i = 1} ^ {16} s_{i}$, and $a = \frac{1}{16}\sum_{i = 1} ^ {16} a_{i}$ for both the analysis period and each 900-second interval. \par
	
	As noted before, in some cases a few of the 16 runs crashed before finishing. They were excluded from the calculation of $m, s$, and $a$ for the analysis period. When calculating $m, s$, and $a$ for the 900-second intervals, data up to the nearest 900-second interval before the run crashed was retained and used for averaging. For example, if a run crashed at 4896 seconds, data up to the 4500 second mark was retained. For the results reported in \autoref{Section3}, if for a case the number of runs that completed successfully is smaller than 10, that case is marked and the number of successful runs is indicated.
	
	\section{Results and Discussion} \label{Section3}
	
	Overall simulation results (traffic delay) for all vehicles during the analysis period are tabulated in \autoref{Table3}. They are divided into blocks based on $q_{i, p}$ (ascending vertically) and $r$ (ascending horizontally). Each horizontal block is further divided by $p_{l}$ values, with baseline being the case where the proposed system was not used. Each vertical block is further divided to show the average, standard deviation of, and maximum delay. The numbers in parenthesis in each row other than the baseline row show percentage change relative to the respective value for the baseline case. For example, the sixth, seventh, and eighth columns of the sixth row show the average, standard deviation of, and maximum delay for the simulation case with $q_{i, p}$ = 4400 veh/hr, $r$ = 40\%, and $p_{l}$ = 0.97, with change relative to the baseline case in the third row indicated by the numbers in parenthesis. Here, average and maximum delay improved by 18.5\% and 9.9\% relative to the baseline case, respectively. Finally, for the few cases where the number of simulations that crashed exceeded 6 (as mentioned before), the number of simulations that did not crash and were used for averaging are indicated as a superscript for the average, standard deviation, and maximum delay values for that case.
	
	\renewcommand{\tabcolsep}{2 pt}
	\begin{table*}[ht!]
		\renewcommand{\arraystretch}{1.2}
		\caption{Statistical characteristics of traffic delay results.} \label{Table3}
		\centering
		\resizebox{\columnwidth}{!}{
		\begin{tabular}{c c c c c c c c c c c c c c}
			\hline
			$q_{i, p}$ & \multirow{2}{*}{$p_{l}$} &  & \multicolumn{3}{c}{$r$ = 10\%} &  & \multicolumn{3}{c}{$r$ = 40\%} &  & \multicolumn{3}{c}{$r$ = 70\%} \\
			\cline{4 - 6} \cline{8 - 10} \cline{12 - 14}
			(veh/hr) &  &  & Avg. (s) & Std. (s) & Max. (s) &  & Avg. (s) & Std. (s) & Max. (s) &  & Avg. (s) & Std. (s) & Max. (s) \\
			\hline
			\multirow{9}{*}{4400} & baseline &  & 29.2 & 33.8 & 153.4 &  & 29.2 & 33.8 & 153.4 &  & 29.2 & 33.8 & 153.4 \\
			 & 0.999 &  & 29.4 (\phantom{-}\phantom{0}0.5) & 33.7 (-\phantom{0}0.3) & 159.9 (\phantom{-}\phantom{0}4.2) &  & 17.8 (-39.1) & 23.3 (-31.0) & 113.3 (-26.1) &  & 46.1 (\phantom{-}57.6) & 53.2 (\phantom{-}57.4) & 219.1 (\phantom{-}42.8) \\
			 & 0.99\phantom{0} &  & 23.8 (-18.8) & 29.1 (-14.0) & 133.2 (-13.2) &  & 21.6 (-26.1) & 26.6 (-21.4) & 131.0 (-14.6) &  & 14.6 (-50.1) & 18.1 (-46.5) & 101.4 (-33.9) \\
			 & 0.97\phantom{0} &  & 30.1 (\phantom{-}\phantom{0}2.9) & 34.9 (\phantom{-}\phantom{0}3.1) & 155.4 (\phantom{-}\phantom{0}1.3) &  & 23.8 (-18.5) & 27.9 (-17.6) & 138.2 (-\phantom{0}9.9) &  & 19.8 (-32.3) & 24.3 (-28.0) & 117.9 (-23.2) \\
			 & 0.95\phantom{0} &  & 23.8 (-18.6) & 30.1 (-11.1) & 143.4 (-\phantom{0}6.5) &  & 21.2 (-27.4) & 25.0 (-25.9) & 129.2 (-15.8) &  & 26.6 (-\phantom{0}9.1) & 30.3 (-10.4) & 138.4 (-\phantom{0}9.8) \\
			 & 0.9\phantom{0}\phantom{0} &  & 33.3 (\phantom{-}13.8) & 37.4 (\phantom{-}10.6) & 165.4 (\phantom{-}\phantom{0}7.8) &  & 20.9 (-28.4) & 24.2 (-28.3) & 117.6 (-23.4) &  & 19.0 (-35.0) & 23.2 (-31.4) & 114.7 (-25.2) \\
			 & 0.85\phantom{0} &  & 29.2 (-\phantom{0}0.2) & 35.4 (\phantom{-}\phantom{0}4.6) & 162.9 (\phantom{-}\phantom{0}6.2) &  & 20.1 (-31.1) & 24.1 (-28.7) & 117.6 (-23.3) &  & 20.6 (-29.4) & 25.8 (-23.7) & 132.5 (-13.7) \\
			 & 0.8\phantom{0}\phantom{0} &  & 29.3 (\phantom{-}\phantom{0}0.2) & 35.7 (\phantom{-}\phantom{0}5.5) & 164.0 (\phantom{-}\phantom{0}6.9) &  & 22.7 (-22.3) & 28.2 (-16.6) & 129.1 (-15.9) &  & 17.5 (-40.2) & 21.9 (-35.3) & 109.9 (-28.4) \\
			 & 0.75\phantom{0} &  & 25.2 (-13.7) & 30.5 (-\phantom{0}9.8) & 141.9 (-\phantom{0}7.5) &  & 27.0 (-\phantom{0}7.6) & 33.1 (-\phantom{0}2.1) & 142.8 (-\phantom{0}6.9) &  & 16.5 (-43.5) & 21.6 (-36.0) & 115.2 (-24.9) \\
			\hline
			\multirow{9}{*}{4600} & baseline &  & 89.2 & 87.0 & 355.1 &  & 89.2 & 87.0 & 355.1 &  & 89.2 & 87.0 & 355.1 \\
			 & 0.999 &  & 78.0 (-12.5) & 80.7 (-7.3) & 305.3 (-14.0) &  & 85.4 (-\phantom{0}4.3) & 86.0 (-\phantom{0}1.2) & 330.8 (-\phantom{0}6.9) &  & 95.2 (\phantom{-}\phantom{0}6.7) & 101.7 (\phantom{-}17.0) & 391.6 (\phantom{-}10.3) \\
			 & 0.99\phantom{0} &  & 90.5 (\phantom{-}\phantom{0}1.5) & 90.5 (\phantom{-}4.0) & 343.7 (-\phantom{0}3.2) &  & 83.7 (-\phantom{0}6.1) & 85.8 (-\phantom{0}1.4) & 327.6 (-\phantom{0}7.7) &  & 85.4 (-\phantom{0}4.2) & \phantom{0}86.5 (-\phantom{0}0.5) & 337.9 (-\phantom{0}4.8) \\
			 & 0.97\phantom{0} &  & 78.1 (-12.5) & 81.5 (-6.3) & 324.3 (-\phantom{0}8.7) &  & 81.3 (-\phantom{0}8.9) & 81.8 (-\phantom{0}5.9) & 342.8 (-\phantom{0}3.5) &  & 71.8 (-19.6) & \phantom{0}74.6 (-14.2) & 302.2 (-14.9) \\
			 & 0.95\phantom{0} &  & 81.4 (-\phantom{0}8.7) & 82.0 (-5.7) & 316.5 (-10.9) &  & 76.2 (-14.5) & 75.8 (-12.8) & 296.9 (-16.4) &  & 69.0 (-22.6) & \phantom{0}77.8 (-10.5) & 312.8 (-11.9) \\
			 & 0.9\phantom{0}\phantom{0} &  & 75.0 (-15.9) & 78.9 (-9.3) & 309.7 (-12.8) &  & 77.7 (-12.9) & 76.7 (-11.8) & 319.1 (-10.1) &  & 81.5 (-\phantom{0}8.6) & \phantom{0}83.9 (-\phantom{0}3.5) & 325.0 (-\phantom{0}8.5) \\
			 & 0.85\phantom{0} &  & 80.0 (-10.3) & 82.5 (-5.1) & 321.2 (-\phantom{0}9.5) &  & 71.3 (-20.1) & 74.8 (-14.0) & 286.8 (-19.2) &  & 84.8 (-\phantom{0}5.0) & \phantom{0}86.3 (-\phantom{0}0.8) & 356.4 (\phantom{-}\phantom{0}0.4) \\
			 & 0.8\phantom{0}\phantom{0} &  & 82.0 (-\phantom{0}8.0) & 86.1 (-1.1) & 333.4 (-\phantom{0}6.1) &  & 73.3 (-17.8) & 75.2 (-13.6) & 295.8 (-16.7) &  & 72.3 (-19.0) & \phantom{0}77.4 (-11.0) & 305.8 (-13.9) \\
			 & 0.75\phantom{0} &  & 80.8 (-\phantom{0}9.4) & 84.6 (-2.7) & 330.6 (-\phantom{0}6.9) &  & 64.2 (-28.0) & 69.9 (-19.7) & 288.9 (-18.7) &  & 80.4 (-\phantom{0}9.9) & \phantom{0}86.3 (-\phantom{0}0.8) & 330.3 (-\phantom{0}7.0) \\
			\hline
			\multirow{9}{*}{4800} & baseline &  & 161.8 & 147.2 & 621.9 &  & 161.8 & 147.2 & 621.9 &  & 161.8 & 147.2 & 621.9 \\
			 & 0.999 &  & 167.8 (\phantom{-}3.7) & 152.4 (\phantom{-}3.6) & 637.5 (\phantom{-}2.5) &  & 158.2 (-\phantom{0}2.2) & 142.8 (-\phantom{0}3.0) & 610.8 (-\phantom{0}1.8) &  & 164.7 (\phantom{-}1.7) \phantom{$^{(8)}$} & 151.9 (\phantom{-}3.2) \phantom{$^{(8)}$} & 669.8 (\phantom{-}7.7) \phantom{$^{(8)}$} \\
			 & 0.99\phantom{0} &  & 161.4 (-0.3) & 146.3 (-0.6) & 623.1 (\phantom{-}0.2) &  & 168.7 (\phantom{-}\phantom{0}4.2) & 151.7 (\phantom{-}\phantom{0}3.1) & 649.4 (\phantom{-}\phantom{0}4.4) &  & 170.6 (\phantom{-}5.4) $^{(8)}$ & 154.6 (\phantom{-}5.1) $^{(8)}$ & 652.2 (\phantom{-}4.9) $^{(8)}$ \\
			 & 0.97\phantom{0} &  & 165.1 (\phantom{-}2.0) & 150.3 (\phantom{-}2.1) & 640.9 (\phantom{-}3.0) &  & 162.2 (\phantom{-}\phantom{0}0.2) & 146.0 (-\phantom{0}0.8) & 632.6 (\phantom{-}\phantom{0}1.7) &  & 172.9 (\phantom{-}6.8) \phantom{$^{(8)}$} & 155.8 (\phantom{-}5.9) \phantom{$^{(8)}$} & 659.6 (\phantom{-}6.1) \phantom{$^{(8)}$} \\
			 & 0.95\phantom{0} &  & 155.2 (-4.1) & 142.8 (-3.0) & 624.7 (\phantom{-}0.5) &  & 166.9 (\phantom{-}\phantom{0}3.2) & 150.3 (\phantom{-}\phantom{0}2.1) & 627.6 (\phantom{-}\phantom{0}0.9) &  & 174.3 (\phantom{-}7.7) \phantom{$^{(8)}$} & 157.1 (\phantom{-}6.7) \phantom{$^{(8)}$} & 674.5 (\phantom{-}8.5) \phantom{$^{(8)}$} \\
			 & 0.9\phantom{0}\phantom{0} &  & 166.9 (\phantom{-}3.1) & 149.4 (\phantom{-}1.6) & 644.5 (\phantom{-}3.6) &  & 163.8 (\phantom{-}\phantom{0}1.2) & 148.9 (\phantom{-}\phantom{0}1.2) & 656.6 (\phantom{-}\phantom{0}5.6) &  & 150.6 (-6.9) \phantom{$^{(8)}$} & 137.1 (-6.8) \phantom{$^{(8)}$} & 589.2 (-5.3) \phantom{$^{(8)}$} \\
			 & 0.85\phantom{0} &  & 161.3 (-0.3) & 146.6 (-0.4) & 621.8 (-0.0) &  & 136.9 (-15.4) & 127.0 (-13.7) & 553.7 (-11.0) &  & 167.0 (\phantom{-}3.2) \phantom{$^{(8)}$} & 152.1 (\phantom{-}3.3) \phantom{$^{(8)}$} & 677.7 (\phantom{-}9.0) \phantom{$^{(8)}$} \\
			 & 0.8\phantom{0}\phantom{0} &  & 158.4 (-2.1) & 144.6 (-1.7) & 623.9 (\phantom{-}0.3) &  & 162.8 (\phantom{-}\phantom{0}0.6) & 148.0 (\phantom{-}\phantom{0}0.6) & 630.4 (\phantom{-}\phantom{0}1.4) &  & 157.7 (-2.5) \phantom{$^{(8)}$} & 146.7 (-0.3) \phantom{$^{(8)}$} & 668.1 (\phantom{-}7.4) \phantom{$^{(8)}$} \\
			 & 0.75\phantom{0} &  & 150.6 (-6.9) & 137.7 (-6.4) & 590.8 (-5.0) &  & 165.3 (\phantom{-}\phantom{0}2.2) & 149.4 (\phantom{-}\phantom{0}1.5) & 638.7 (\phantom{-}\phantom{0}2.7) &  & 158.7 (-1.9) $^{(9)}$ & 146.1 (-0.7) $^{(9)}$ & 609.4 (-2.0) $^{(9)}$ \\
			\hline
		\end{tabular}}
	\end{table*}
	\renewcommand{\tabcolsep}{6 pt}
	
	An overall look at the results shows that in all cases (combinations of $q_{i, p}$ and $r$), for at least one value of $p_{l}$ the system was successful at reducing average traffic delay, but its behavior varied for different cases. Therefore, after discussing some broad trends in the results, we look at two cases in more detail: the case with $q_{i, p}$ = 4600 veh/hr and $r$ = 40\% where all values of $p_{l}$ result in a sizable reduction in delay; and the case with $q_{i, p}$ = 4800 veh/hr and $r$ = 70\% where only a few $p_{l}$ values result in minor reductions in delay. From here on, the former is called Case A and the latter Case B. \par
	
	\autoref{Table3} shows that as $q_{i, p}$ increases, so does average delay for the baseline case. This is because as vehicle density per lane increases, for vehicles traveling in the blocked lanes finding an acceptable gap in an adjacent lane becomes harder and changing lanes results in additional traffic disruptions and delays. In the same light, the results show that as $q_{i, p}$ increases and the road becomes congested beyond the capacity of a two-lane highway, the impact of the system is reduced. For baseline cases, as the advance warning system is not active its penetration rate $r$ does not have an impact on the results. When it is active, for $r$ = 10\% the results match our expectations that regardless of $q_{i, p}$, because of a low penetration rate the system would only have a modest impact on delay. For $q_{i, p}$ = 4400 veh/hr, reduction in delay is largest when $r$ = 70\%, with improvements reaching as high as 50\%. This is reversed, however, for the other two values of $q_{i, p}$, because for higher penetration rates the system tries to force more traffic onto non-blocked lanes earlier, causing additional delay. Finally, broadly speaking to the effects of $p_{l}$, when it is higher (for example 0.999) the system warns drivers much earlier than it would when $p_{l}$ is lower, which depending on $q_{i, p}$ and $r$ can have a positive or negative impact on traffic, as will be discussed next. For all parameters, similar trends can be observed for maximum and standard deviation values of delay. \par
	
	\addtocounter{equation}{2}
	
	Before moving forward, we need to define a parameter called lane departure density denoted by $d_{l}$. For a blocked lane, if $N$ vehicles depart that lane for the final time in a specific road span $D$ and time span $T$ during the simulation, $d_{l}$ for that $T$-$D$ time-space block is defined as
	\begin{equation} \label{Equation4}
		d_{l} = \frac{N}{\vert D \vert\vert T \vert},
	\end{equation}
	with (lane departure)/(ft.s) as its unit. In other words, $d_{l}$ quantifies the time-space rate of vehicles leaving a blocked lane for the last time during the simulation. For simplicity, in this paper we assume that $D$ = 100 ft and $T$ = 100 s. \par
	
	Detailed average delay results for Case A and Case B are shown in \autoref{Table4} and \autoref{Table5}, respectively. In both tables, time intervals are shown in the first column and the results for the baseline case are shown in the second column. Starting from the third column, each one presents the results for a $p_{l}$ value, starting from $p_{l}$ = 0.999 in descending order. Each row represents a 900-second simulation time interval during the analysis period. As before, numbers in parenthesis show change in average delay relative to the baseline case of the respective time interval.
	
	\renewcommand{\tabcolsep}{4 pt}
	\begin{table*}[ht!]
		\renewcommand{\arraystretch}{1.2}
		\caption{Average delay for $q_{i, p}$ = 4600 veh/hr and $r$ = 40\% (Case A).} \label{Table4}
		\centering
		\resizebox{\columnwidth}{!}{
		\begin{tabular}{c c c c c c c c c c}
			\hline
			Time & \multicolumn{9}{c}{$p_{l}$} \\
			\cline{2 - 10}
			Interval (s) & baseline & 0.999 & 0.99 & 0.97 & 0.95 & 0.9 & 0.85 & 0.8 & 0.75 \\
			\hline
			1800 - 2700 & \phantom{0}\phantom{0}1.5 & \phantom{0}\phantom{0}1.5 (\phantom{-}\phantom{0}1.9) & \phantom{0}\phantom{0}1.5 (\phantom{-}\phantom{0}1.4) & \phantom{0}\phantom{0}1.5 (\phantom{-}\phantom{0}0.2) & \phantom{0}\phantom{0}1.5 (-\phantom{0}0.0) & \phantom{0}\phantom{0}1.5 (-\phantom{0}0.1) & \phantom{0}\phantom{0}1.5 (-\phantom{0}0.3) & \phantom{0}\phantom{0}1.5 (-\phantom{0}0.4) & \phantom{0}\phantom{0}1.5 (-\phantom{0}0.6) \\
			2700 - 3600 & \phantom{0}\phantom{0}1.6 & \phantom{0}\phantom{0}1.6 (\phantom{-}\phantom{0}1.3) & \phantom{0}\phantom{0}1.6 (\phantom{-}\phantom{0}1.3) & \phantom{0}\phantom{0}1.6 (\phantom{-}\phantom{0}1.6) & \phantom{0}\phantom{0}1.6 (\phantom{-}\phantom{0}0.9) & \phantom{0}\phantom{0}1.6 (\phantom{-}\phantom{0}1.0) & \phantom{0}\phantom{0}1.6 (\phantom{-}\phantom{0}0.7) & \phantom{0}\phantom{0}1.6 (\phantom{-}\phantom{0}0.3) & \phantom{0}\phantom{0}1.6 (\phantom{-}\phantom{0}0.1) \\
			3600 - 4500 & \phantom{0}16.8 & \phantom{0}14.7 (-12.7) & \phantom{0}11.7 (-30.0) & \phantom{0}13.2 (-21.2) & \phantom{0}13.2 (-21.4) & \phantom{0}14.9 (-11.2) & \phantom{0}11.6 (-30.8) & \phantom{0}11.1 (-34.0) & \phantom{0}\phantom{0}9.4 (-43.7) \\
			4500 - 5400 & \phantom{0}85.4 & \phantom{0}82.9 (-\phantom{0}3.0) & \phantom{0}76.4 (-10.5) & \phantom{0}72.9 (-14.7) & \phantom{0}73.8 (-13.6) & \phantom{0}73.0 (-14.5) & \phantom{0}63.7 (-25.4) & \phantom{0}67.1 (-21.5) & \phantom{0}50.2 (-41.3) \\
			5400 - 6300 & 141.8 & 138.5 (-\phantom{0}2.3) & 137.4 (-\phantom{0}3.1) & 128.3 (-\phantom{0}9.5) & 122.1 (-13.9) & 127.1 (-10.4) & 120.6 (-15.0) & 121.5 (-14.3) & 102.9 (-27.4) \\
			6300 - 7200 & 190.0 & 187.1 (-\phantom{0}1.5) & 189.2 (-\phantom{0}0.4) & 173.4 (-\phantom{0}8.8) & 167.4 (-11.9) & 163.6 (-13.9) & 161.7 (-14.9) & 163.1 (-14.2) & 150.2 (-20.9) \\
			7200 - 8100 & 162.6 & 150.7 (-\phantom{0}7.3) & 145.4 (-10.5) & 154.3 (-\phantom{0}5.1) & 126.3 (-22.3) & 137.1 (-15.7) & 122.3 (-24.8) & 128.8 (-20.8) & 117.2 (-27.9) \\
			8100 - 9000 & \phantom{0}\phantom{0}2.8 & \phantom{0}\phantom{0}2.7 (-\phantom{0}3.9) & \phantom{0}\phantom{0}2.0 (-30.1) & \phantom{0}\phantom{0}2.1 (-26.7) & \phantom{0}\phantom{0}1.5 (-48.2) & \phantom{0}\phantom{0}2.4 (-14.3) & \phantom{0}\phantom{0}1.5 (-47.3) & \phantom{0}\phantom{0}1.5 (-47.2) & \phantom{0}\phantom{0}1.5 (-47.4) \\
			\hline
		\end{tabular}}
	\end{table*}
	\renewcommand{\tabcolsep}{6 pt}
	\renewcommand{\tabcolsep}{5 pt}
	\begin{table*}[ht!]
		\renewcommand{\arraystretch}{1.2}
		\caption{Average delay for $q_{i, p}$ = 4800 veh/hr and $r$ = 70\% (Case B).} \label{Table5}
		\centering
		\resizebox{\columnwidth}{!}{
		\begin{tabular}{c c c c c c c c c c}
			\hline
			Time & \multicolumn{9}{c}{$p_{l}$} \\
			\cline{2 - 10}
			Interval (s) & baseline & 0.999 & 0.99 & 0.97 & 0.95 & 0.9 & 0.85 & 0.8 & 0.75 \\
			\hline
			1800 - 2700 & \phantom{0}\phantom{0}1.5 & \phantom{0}\phantom{0}1.6 (\phantom{-}\phantom{0}7.2) & \phantom{0}\phantom{0}1.5 (\phantom{-}\phantom{0}4.5) & \phantom{0}\phantom{0}1.5 (\phantom{-}\phantom{0}1.5) & \phantom{0}\phantom{0}1.5 (-\phantom{0}0.5) & \phantom{0}\phantom{0}1.5 (-0.4) & \phantom{0}\phantom{0}1.5 (-\phantom{0}0.4) & \phantom{0}\phantom{0}1.5 (-\phantom{0}0.7) & \phantom{0}\phantom{0}1.5 (-1.0) \\
			2700 - 3600 & \phantom{0}\phantom{0}1.6 & \phantom{0}\phantom{0}1.7 (\phantom{-}\phantom{0}6.7) & \phantom{0}\phantom{0}1.6 (\phantom{-}\phantom{0}4.9) & \phantom{0}\phantom{0}1.6 (\phantom{-}\phantom{0}1.3) & \phantom{0}\phantom{0}1.6 (\phantom{-}\phantom{0}0.1) & \phantom{0}\phantom{0}1.6 (\phantom{-}0.3) & \phantom{0}\phantom{0}1.6 (-\phantom{0}0.0) & \phantom{0}\phantom{0}1.6 (-\phantom{0}0.2) & \phantom{0}\phantom{0}1.6 (-0.7) \\
			3600 - 4500 & \phantom{0}30.9 & \phantom{0}26.6 (-13.9) & \phantom{0}26.0 (-15.7) & \phantom{0}30.0 (-\phantom{0}2.9) & \phantom{0}33.5 (\phantom{-}\phantom{0}8.6) & \phantom{0}29.7 (-3.9) & \phantom{0}32.5 (\phantom{-}\phantom{0}5.4) & \phantom{0}30.3 (-\phantom{0}1.9) & \phantom{0}28.8 (-6.7) \\
			4500 - 5400 & 140.5 & 138.5 (-\phantom{0}1.4) & 147.6 (\phantom{-}\phantom{0}5.0) & 150.4 (\phantom{-}\phantom{0}7.1) & 151.7 (\phantom{-}\phantom{0}8.0) & 132.1 (-6.0) & 141.8 (\phantom{-}\phantom{0}0.9) & 137.7 (-\phantom{0}2.0) & 130.8 (-6.9) \\
			5400 - 6300 & 229.1 & 225.2 (-\phantom{0}1.7) & 246.2 (\phantom{-}\phantom{0}7.4) & 227.7 (-\phantom{0}0.6) & 242.2 (\phantom{-}\phantom{0}5.7) & 216.8 (-5.4) & 228.2 (-\phantom{0}0.4) & 225.4 (-\phantom{0}1.6) & 215.5 (-6.0) \\
			6300 - 7200 & 316.9 & 317.5 (\phantom{-}\phantom{0}0.2) & 325.6 (\phantom{-}\phantom{0}2.8) & 333.9 (\phantom{-}\phantom{0}5.4) & 334.9 (\phantom{-}\phantom{0}5.7) & 295.3 (-6.8) & 321.3 (\phantom{-}\phantom{0}1.4) & 312.2 (-\phantom{0}1.5) & 308.0 (-2.8) \\
			7200 - 8100 & 347.5 & 362.6 (\phantom{-}\phantom{0}4.4) & 368.3 (\phantom{-}\phantom{0}6.0) & 378.1 (\phantom{-}\phantom{0}8.8) & 373.6 (\phantom{-}\phantom{0}7.5) & 324.3 (-6.7) & 368.7 (\phantom{-}\phantom{0}6.1) & 356.8 (\phantom{-}\phantom{0}2.7) & 340.7 (-2.0) \\
			8100 - 9000 & \phantom{0}36.6 & \phantom{0}53.8 (\phantom{-}47.0) & \phantom{0}65.3 (\phantom{-}78.3) & \phantom{0}61.9 (\phantom{-}69.1) & \phantom{0}55.0 (\phantom{-}50.3) & \phantom{0}33.9 (-7.4) & \phantom{0}44.8 (\phantom{-}22.4) & \phantom{0}46.1 (\phantom{-}26.0) & \phantom{0}35.5 (-3.1) \\
			\hline
		\end{tabular}}
	\end{table*}
	\renewcommand{\tabcolsep}{6 pt}
	
	\begin{figure*}[t!]
		\centering
		\includegraphics[width = 0.27\columnwidth]{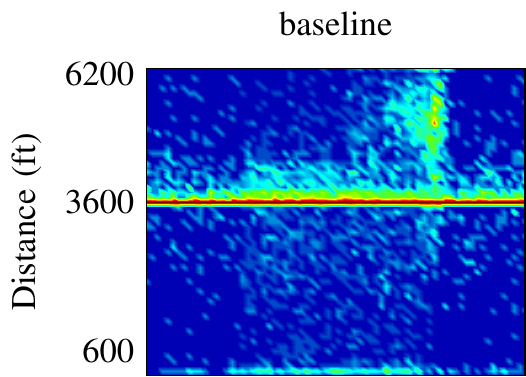} \label{Figure4P1}
		\includegraphics[width = 0.196\columnwidth]{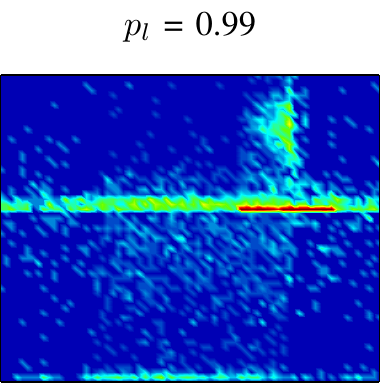} \label{Figure4P2}
		\includegraphics[width = 0.196\columnwidth]{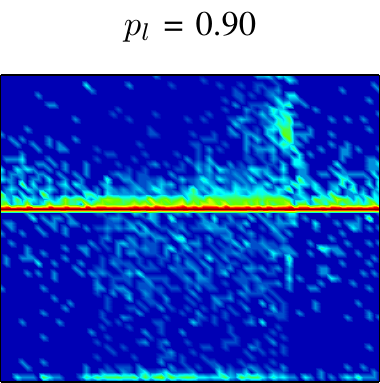} \label{Figure4P3}
		\includegraphics[width = 0.224\columnwidth]{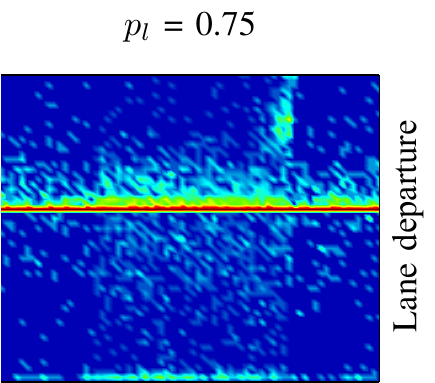} \label{Figure4P4}
		\includegraphics[width = 0.27\columnwidth]{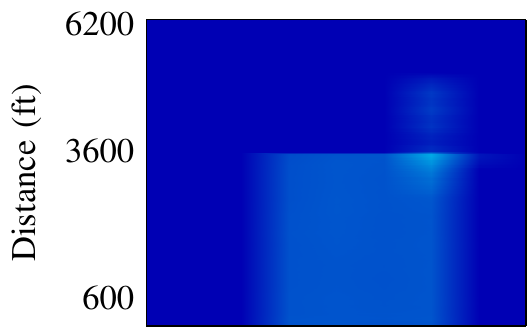} \label{Figure4P5}
		\includegraphics[width = 0.196\columnwidth]{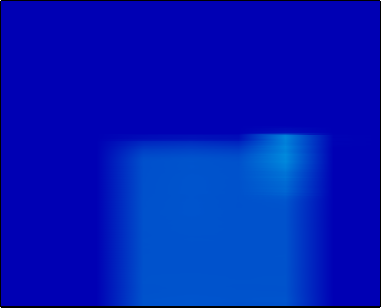} \label{Figure4P6}
		\includegraphics[width = 0.196\columnwidth]{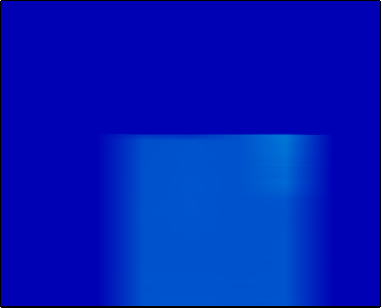} \label{Figure4P7}
		\includegraphics[width = 0.224\columnwidth]{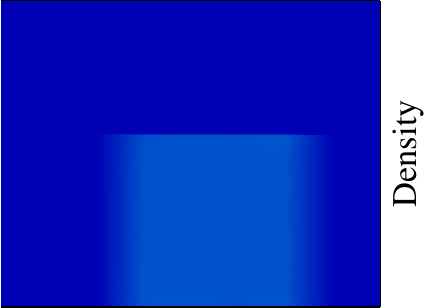} \label{Figure4P8}
		\includegraphics[width = 0.27\columnwidth]{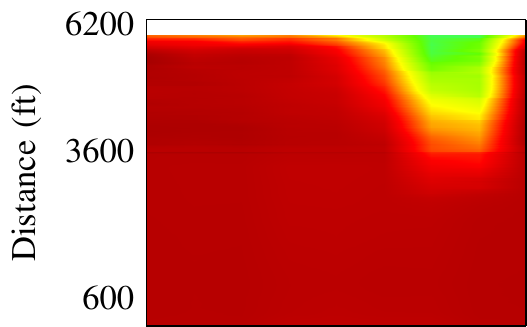} \label{Figure4P9}
		\includegraphics[width = 0.196\columnwidth]{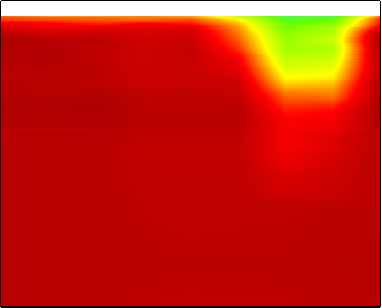} \label{Figure4P10}
		\includegraphics[width = 0.196\columnwidth]{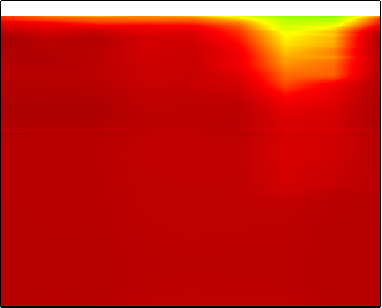} \label{Figure4P11}
		\includegraphics[width = 0.224\columnwidth]{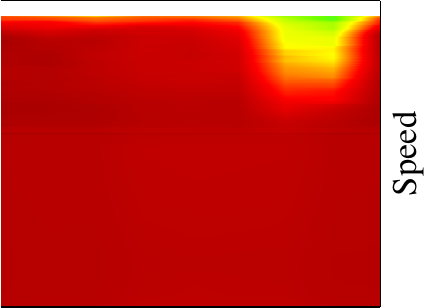} \label{Figure4P12}
		\includegraphics[width = 0.27\columnwidth]{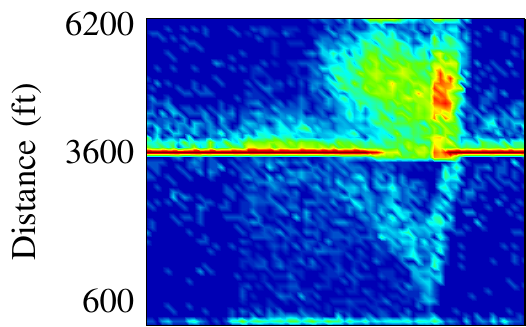} \label{Figure4P13}
		\includegraphics[width = 0.196\columnwidth]{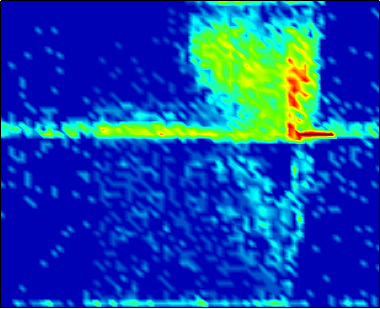} \label{Figure4P14}
		\includegraphics[width = 0.196\columnwidth]{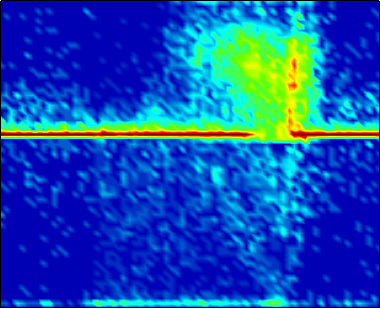} \label{Figure4P15}
		\includegraphics[width = 0.224\columnwidth]{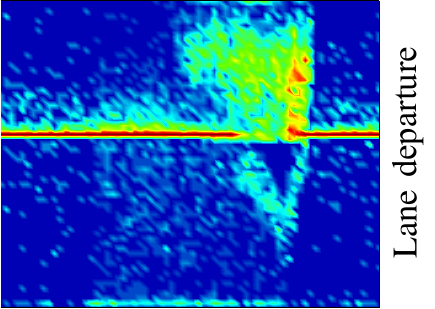} \label{Figure4P16}
		\includegraphics[width = 0.27\columnwidth]{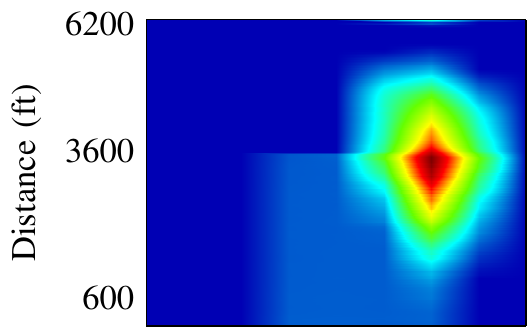} \label{Figure4P17}
		\includegraphics[width = 0.196\columnwidth]{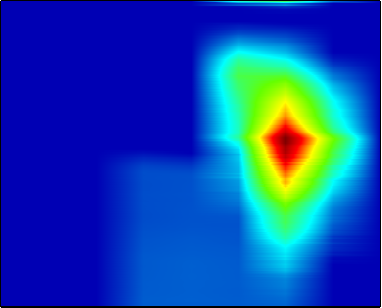} \label{Figure4P18}
		\includegraphics[width = 0.196\columnwidth]{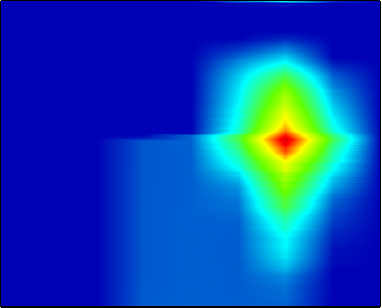} \label{Figure4P19}
		\includegraphics[width = 0.224\columnwidth]{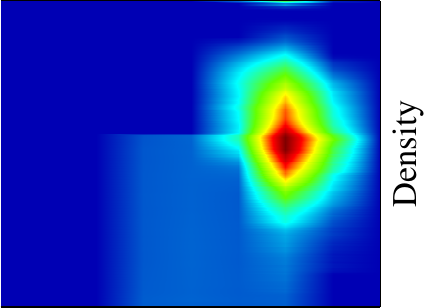} \label{Figure4P20}
		\includegraphics[width = 0.27\columnwidth]{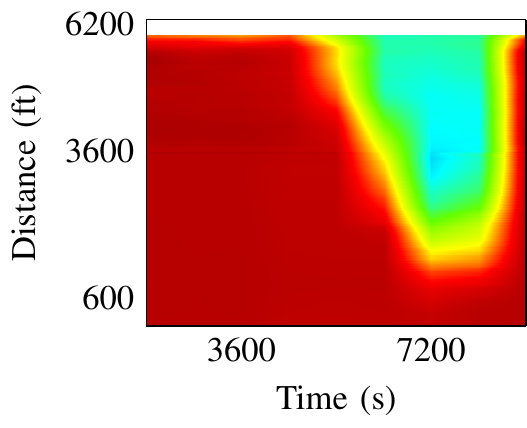} \label{Figure4P21}
		\includegraphics[width = 0.196\columnwidth]{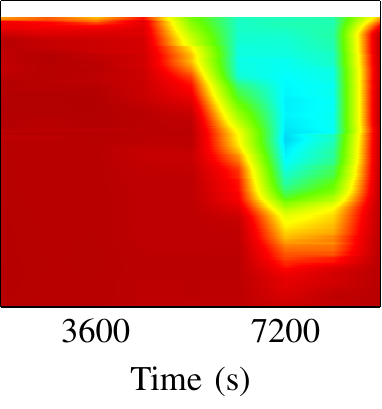} \label{Figure4P22}
		\includegraphics[width = 0.196\columnwidth]{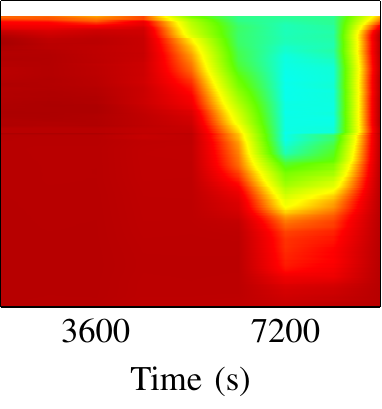} \label{Figure4P23}
		\includegraphics[width = 0.224\columnwidth]{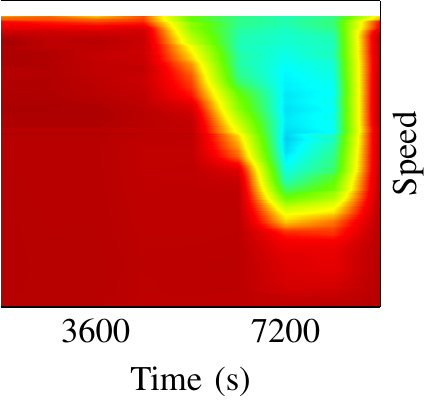} \label{Figure4P24}
		\includegraphics[width = 0.3\columnwidth]{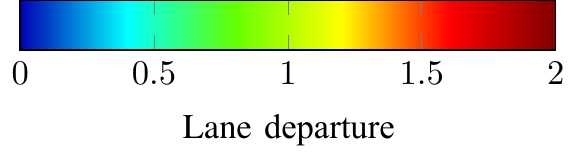} \label{Figure4P25}
		\includegraphics[width = 0.3\columnwidth]{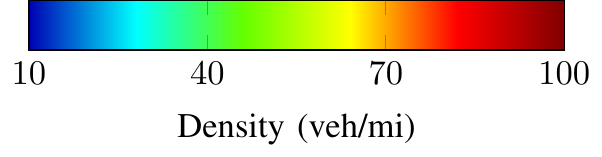} \label{Figure4P26}
		\includegraphics[width = 0.3\columnwidth]{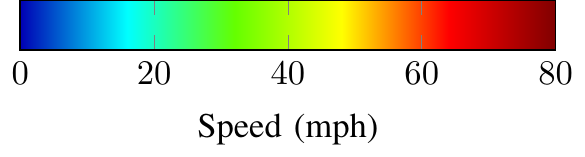} \label{Figure4P27}
		\caption{Time-space plots of density and speed of all vehicles and lane departure of smart cars for the fourth (leftmost) lane. The top half plots belong to Case A and the bottom half to Case B.} \label{Figure4}
	\end{figure*}
	\begin{figure*}[t!]
		\centering
		\includegraphics[width = 0.27\columnwidth]{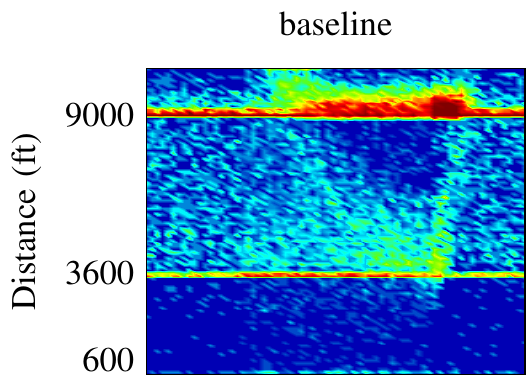} \label{Figure5P1}
		\includegraphics[width = 0.196\columnwidth]{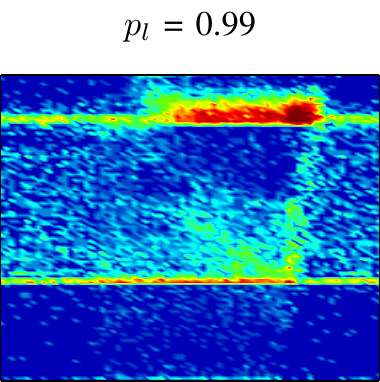} \label{Figure5P2}
		\includegraphics[width = 0.196\columnwidth]{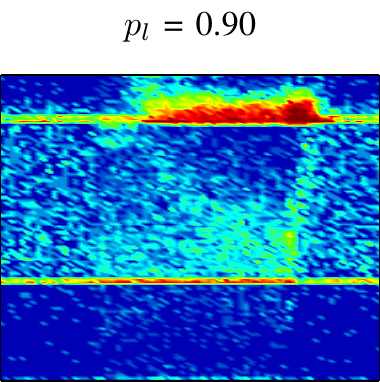} \label{Figure5P3}
		\includegraphics[width = 0.224\columnwidth]{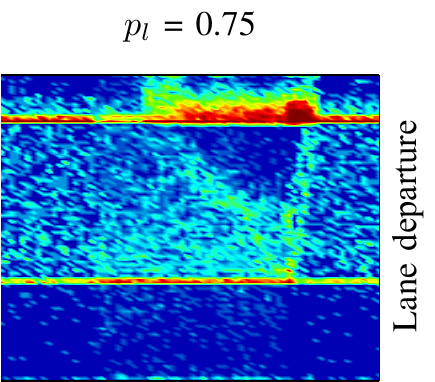} \label{Figure5P4}
		\vspace*{4 pt}
		\includegraphics[width = 0.27\columnwidth]{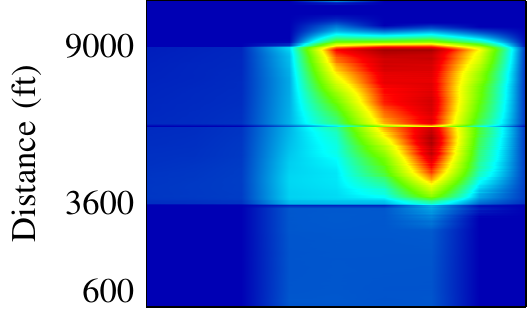} \label{Figure5P5}
		\includegraphics[width = 0.196\columnwidth]{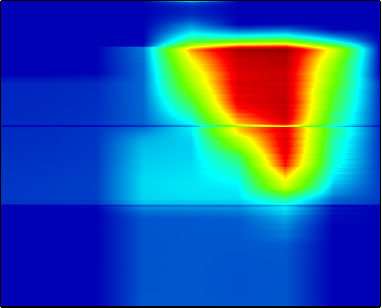} \label{Figure5P6}
		\includegraphics[width = 0.196\columnwidth]{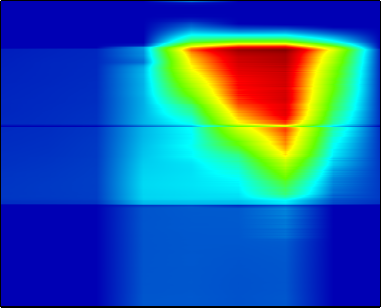} \label{Figure5P7}
		\includegraphics[width = 0.224\columnwidth]{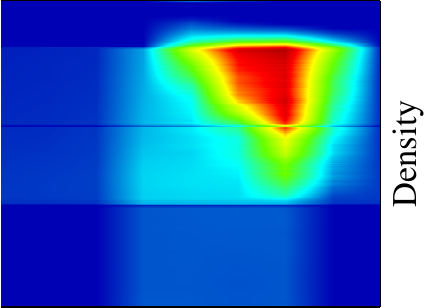} \label{Figure5P8}
		\vspace*{4 pt}
		\includegraphics[width = 0.27\columnwidth]{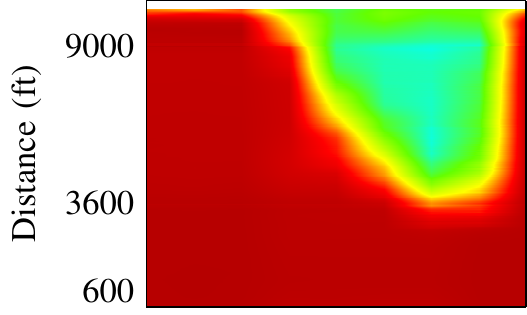} \label{Figure5P9}
		\includegraphics[width = 0.196\columnwidth]{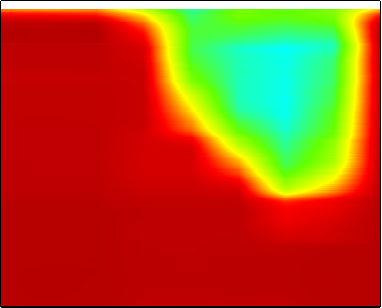} \label{Figure5P10}
		\includegraphics[width = 0.196\columnwidth]{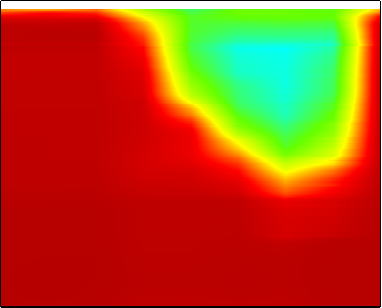} \label{Figure5P11}
		\includegraphics[width = 0.224\columnwidth]{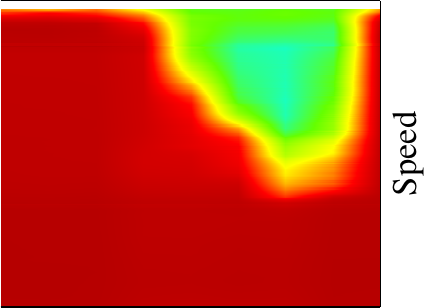} \label{Figure5P12}
		\vspace*{4 pt}
		\includegraphics[width = 0.27\columnwidth]{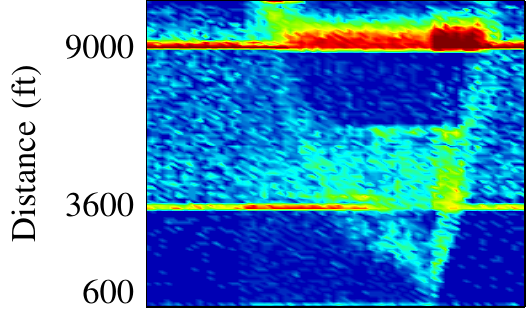} \label{Figure5P13}
		\includegraphics[width = 0.196\columnwidth]{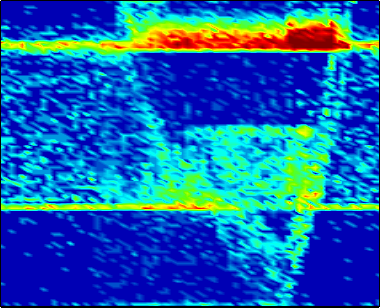} \label{Figure5P14}
		\includegraphics[width = 0.196\columnwidth]{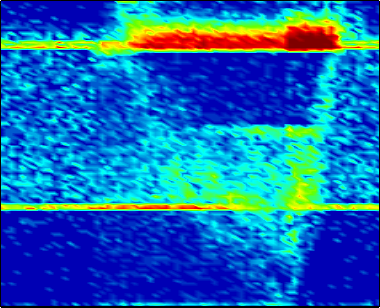} \label{Figure5P15}
		\includegraphics[width = 0.224\columnwidth]{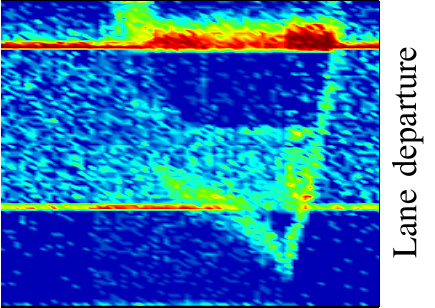} \label{Figure5P16}
		\vspace*{4 pt}
		\includegraphics[width = 0.27\columnwidth]{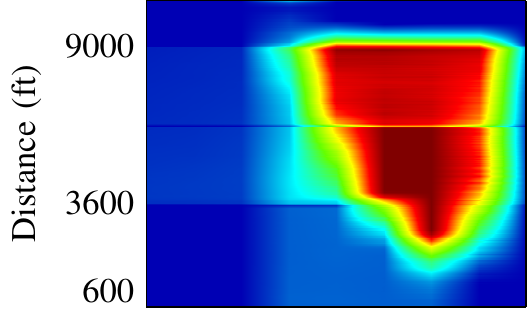} \label{Figure5P17}
		\includegraphics[width = 0.196\columnwidth]{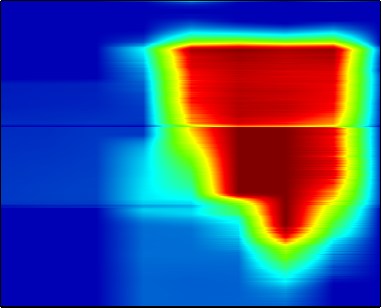} \label{Figure5P18}
		\includegraphics[width = 0.196\columnwidth]{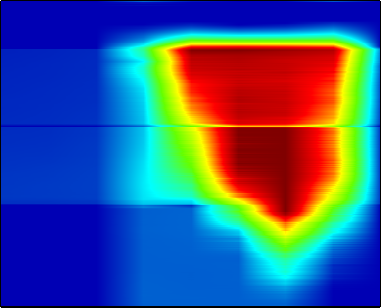} \label{Figure5P19}
		\includegraphics[width = 0.224\columnwidth]{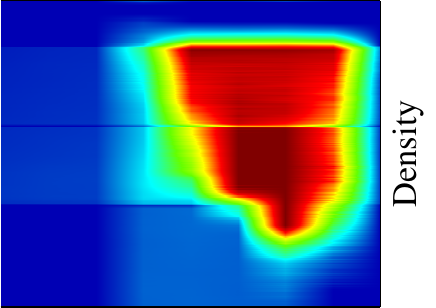} \label{Figure5P20}
		\vspace*{4 pt}
		\hspace*{-6 pt}
		\includegraphics[width = 0.27\columnwidth]{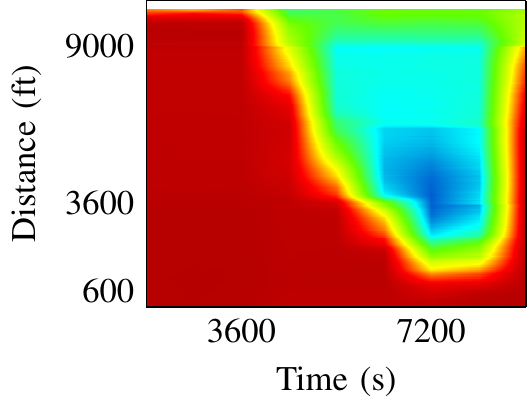} \label{Figure5P21}
		\includegraphics[width = 0.196\columnwidth]{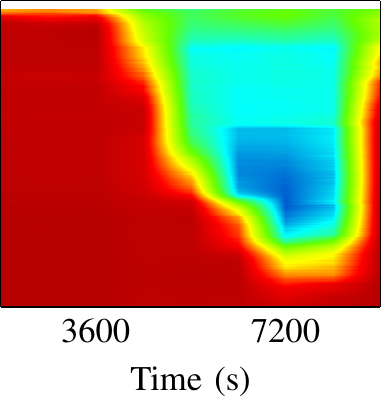} \label{Figure5P22}
		\includegraphics[width = 0.196\columnwidth]{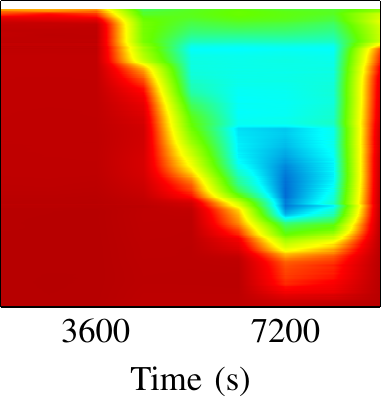} \label{Figure5P23}
		\includegraphics[width = 0.224\columnwidth]{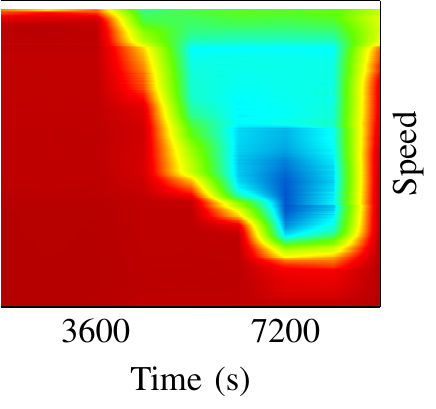} \label{Figure5P24}
		\includegraphics[width = 0.3\columnwidth]{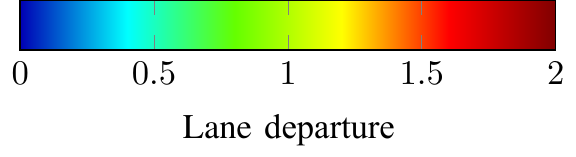} \label{Figure5P25}
		\includegraphics[width = 0.3\columnwidth]{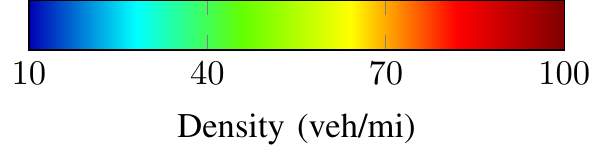} \label{Figure5P26}
		\includegraphics[width = 0.3\columnwidth]{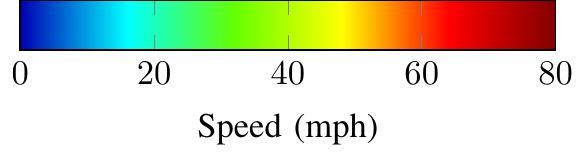} \label{Figure5P27}
		\caption{Time-space plots of density and speed of all vehicles and lane departure of smart cars for the third lane (from right). The top half plots belong to Case A and the bottom half to Case B.} \label{Figure5}
	\end{figure*}
		\begin{figure*}[t!]
		\centering
		\includegraphics[width = 0.28\columnwidth]{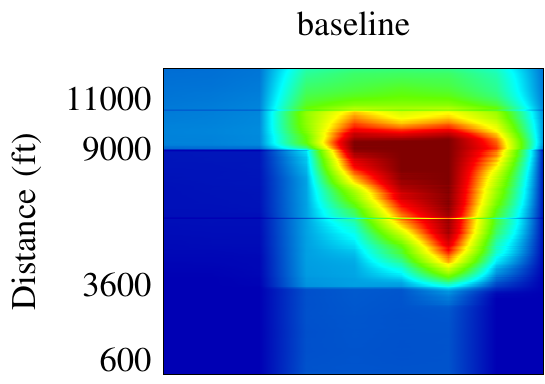} \label{Figure6P1}
		\includegraphics[width = 0.196\columnwidth]{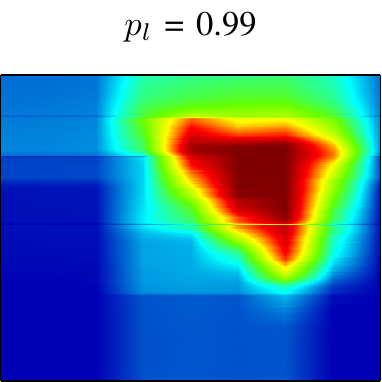} \label{Figure6P2}
		\includegraphics[width = 0.196\columnwidth]{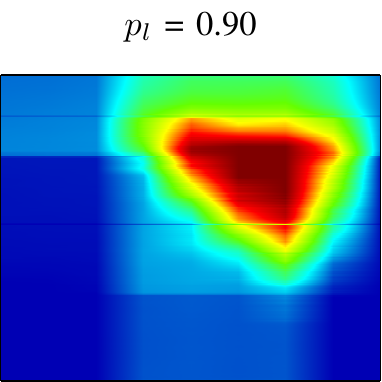} \label{Figure6P3}
		\includegraphics[width = 0.224\columnwidth]{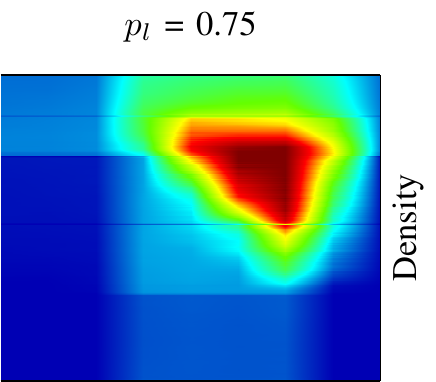} \label{Figure6P4}
		\vspace*{4 pt}
		\includegraphics[width = 0.28\columnwidth]{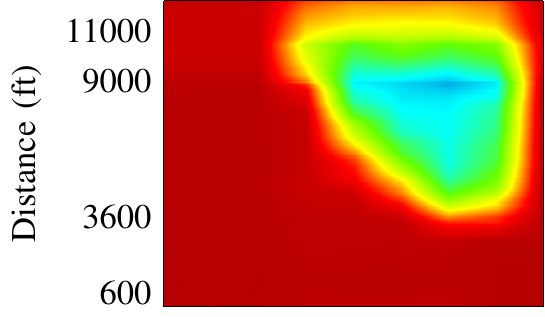} \label{Figure6P5}
		\includegraphics[width = 0.196\columnwidth]{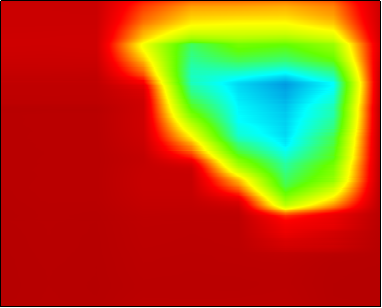} \label{Figure6P6}
		\includegraphics[width = 0.196\columnwidth]{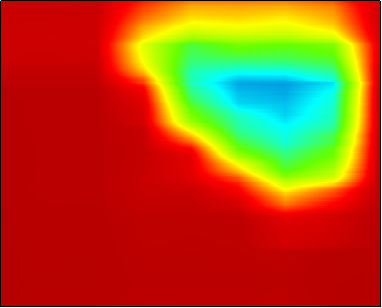} \label{Figure6P7}
		\includegraphics[width = 0.224\columnwidth]{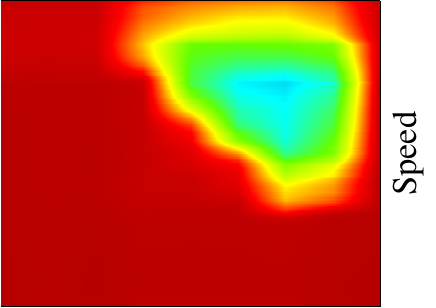} \label{Figure6P8}
		\vspace*{4 pt}
		\includegraphics[width = 0.28\columnwidth]{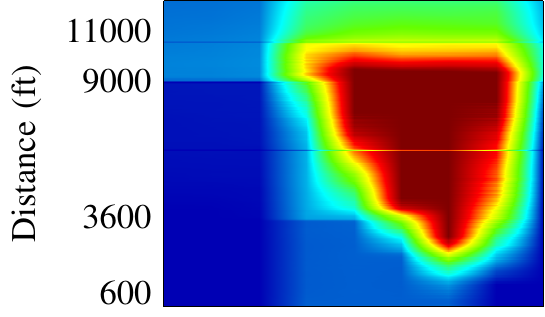} \label{Figure6P9}
		\includegraphics[width = 0.196\columnwidth]{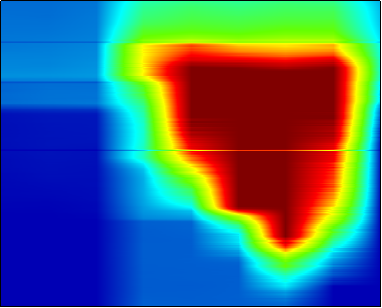} \label{Figure6P10}
		\includegraphics[width = 0.196\columnwidth]{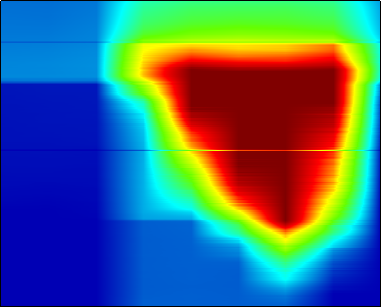} \label{Figure6P11}
		\includegraphics[width = 0.224\columnwidth]{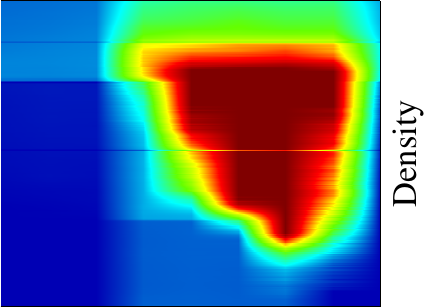} \label{Figure6P12}
		\vspace*{4 pt}
		\includegraphics[width = 0.28\columnwidth]{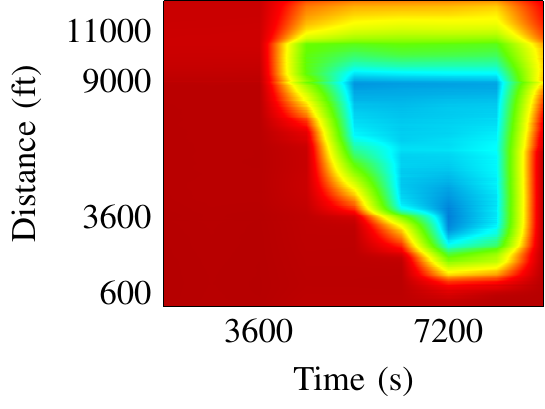} \label{Figure6P13}
		\includegraphics[width = 0.196\columnwidth]{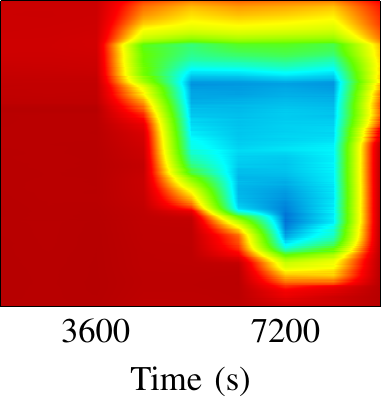} \label{Figure6P14}
		\includegraphics[width = 0.196\columnwidth]{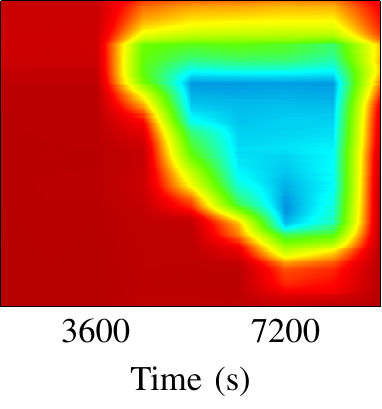} \label{Figure6P15}
		\includegraphics[width = 0.224\columnwidth]{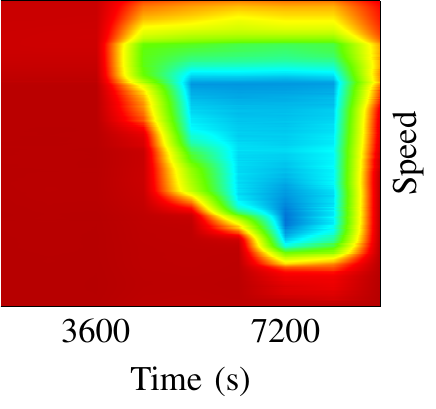} \label{Figure6P16}
		\vspace*{4 pt}
		\includegraphics[width = 0.3\columnwidth]{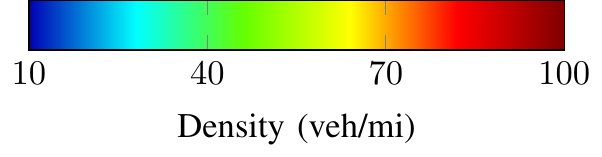} \label{Figure6P17}
		\includegraphics[width = 0.3\columnwidth]{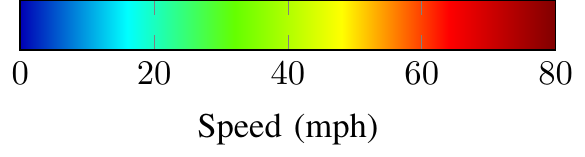} \label{Figure6P18}
		\caption{Time-space plots of density and speed of all vehicles for the second lane (from right). The top half plots belong to Case A and the bottom half to Case B.} \label{Figure6}
	\end{figure*}
	
	To analyze the performance of the proposed system, time-space plots of density (in veh/mi) and speed (in mph) for all vehicles and $\log(K\frac{d_{l}}{q_{i}r} + 1)$ for smart cars are shown in \autoref{Figure4} for lane 4 and in \autoref{Figure5} for lane 3. Time-space plots of density and speed for lane 2 are shown in \autoref{Figure6}. In each figure, the top half plots belong to Case A and the bottom half to Case B. For each case, plot rows represent lane departure, density, and speed, while plot columns represent the baseline case and cases with $p_{l}$ = 0.99, 0.9, and 0.75, respectively. For better referencing, plot rows of each figure are assigned a letter from A to F from top to bottom (A to D in the case of \autoref{Figure6}) and plot columns are assigned a number from 1 to 4 from left to right. For example, \autoref{Figure5}C3 refers to the plot on the third row and third column of \autoref{Figure5}, showing time-space variation of speed for a simulation of Case A where $p_{l}$ = 0.9. \par
	
	The time axis of each plot spans the entire analysis period (1800 to 9000 seconds) and the distance axis spans the entire distance of the respective lane, i.e. 6277 ft for the fourth lane, 10575 ft for the third lane, and 12208 ft for the second lane. Furthermore, $\log(K\frac{d_{l}}{q_{i}r} + 1)$ was plotted instead of $d_{l}$ to i) make the parameter independent of $q_{i}$ (in units of veh/s) and $r$, ii) simplify presentation by using a constant scaling factor $K$ = 10000, and iii) better display differences between cases by using logarithmic values. Finally, note that while speed and density plots are made using data from all vehicles, lane departure plots are made using data from only smart cars. This is because turning the proposed advance warning system off (as in the baseline case) or on (as in other cases) only affects the behavior of smart cars and the behavior of other cars should not be statistically different from that shown for the baseline case where the system is turned off. \par
	
	Lane departure plots of both Cases A and B in \autoref{Figure4} show a narrow horizontal band at around 3600 ft where lane departure from the fourth lane peaks. That area corresponds to the road segment just after the lane-end sign for the first lane drop and shows that a large portion of smart cars (and other cars) leave that lane after seeing the sign, conditioned on traffic density and speed of the adjacent lane. Similar bands can be seen in lane departure plots of \autoref{Figure5}, this time at around 3600 ft and 9000 ft. Same as before, the latter corresponds to the road segment just after the lane-end sign for the second lane drop. As for the former, it indicates cars that move from the third lane to the second lane to make room for vehicles that are coming from the fourth lane. \par
	
	To understand the effect of $p_{l}$ on the change in average delay, we first take a look at Case A. As previously mentioned, when $p_{l}$ is higher the system warns drivers much earlier than it would when $p_{l}$ is lower. This can be seen when comparing \autoref{Figure5}A2 to \autoref{Figure5}A4. When $p_{l}$ is 0.99, lane departures happen much earlier than when it is 0.9 or 0.75, as indicated by a smaller red area near the 9000 ft band. This initially helps reduce average delay during the peak period by pushing some lane departures away from the concentration at the 9000 ft band, as evidenced from \autoref{Table4} where average delay for the 3600 - 4500-second time interval is reduced by 30\% compared to the baseline where $p_{l}$ is 0.99. However, because peak traffic flow is near the capacity of a two-lane highway, the congestion at the second lane drop eventually grows and this pushes the point where the system warns drivers further and further back, until it coincides with the 3600 ft band where vehicles are already moving from the third lane to the second lane because of those moving from the fourth lane to the third lane. This increased volume of vehicles moving to the second lane causes congestion and reduces speed in that lane, as evidenced by \autoref{Figure6}A2 and \autoref{Figure6}B2, increasing the delay. In comparison, $p_{l}$ = 0.9 and 0.75 do a better job of distributing lane departures along the segment of the third lane between the 3600 ft and 9000 ft bands, reducing average delay by more in the end even though they may have lagged initially. In comparison to the baseline case, all three cases are successful in reducing average delay because they significantly delay or slow the growth of congestion at the second lane drop. \par
	
	A similar story plays out for Case B. Same as before, when $p_{l}$ is 0.99, lane departures happen much earlier than when it is 0.9 or 0.75, as evidenced by the smaller red area near the 9000 ft band of \autoref{Figure5}D2. By pushing some lane departures away from the concentration at the 9000 ft band, the system helps reduce average delay during the 3600 - 4500-second time interval by around 16\%, much larger than the 4\% reduction for $p_{l}$ = 0.9 or the 7\% reduction for $p_{l}$ = 0.75.  As in Case A, the congestion eventually grows and pushes the point of warning further and further back until it reaches the 3600 ft band. Denser traffic and higher ratio of smart cars compared to Case A combined with the volume of vehicles already departing the third lane to make room for those departing the fourth lane causes a big surge in the density of the second lane and slows traffic down. This can be seen in \autoref{Figure6}C2 and \autoref{Figure6}C3, where the boundary of the red area is much steeper in the former than it is in the latter. Between $p_{l}$ values of 0.75 and 0.9, for the former the warning to change lanes comes too late and too close to the 9000 ft band, while the latter does the best overall job of distributing lane departures to balance the increase in density in the second lane. Compared to the baseline case, $p_{l}$ = 0.99 increases average delay by 5\% while the other two decrease it, though not by much when $p_{l}$ is 0.75. \par
	
	So how should we select $p_{l}$ for a different case? As discussed above, the answer depends on traffic flow and system penetration rate, but given that larger values of $p_{l}$ tend to push some lane departures away from concentration zones after lane-end signs and distribute them more evenly, a rule of thumb for general cases (involving one lane drop) would be to use larger values in the range of 0.9 to 0.99. For other, more complex cases like the one discussed here, the answer may require additional traffic simulation. Another possible solution may be to dynamically assign $p_{l}$ based on various traffic flow characteristics, though this strategy needs further research. \par
	
	Compared to other methods, the proposed system has two main advantages. The first is that it can be implemented in a simple, cost-effective way. A real-world implementation would only need traffic information, distance to the lane drop, and vehicle velocity to calculate the probability. The first one can be obtained from real-time or existing traffic data, possibly stored as a database where the system can search based on vehicle location and time of day. The second one can be calculated based on the position of the vehicle, and the last one can be obtained directly from the vehicle. This means the proposed system can be directly integrated with the in-vehicle navigation system. The second advantage is that it can be used together with other delay reduction strategies such as VSLs, as it only affects the lane changing behavior of vehicles and not their longitudinal behavior, though more research is needed in this area.
	
	\section{Conclusions and Outlook} \label{Section4}
	
	In this work we proposed an onboard advance warning system based on a probabilistic prediction model that advised vehicles on when to change lanes for an upcoming lane drop. Using a variety of traffic- and driver- related parameters such as the distribution of inter-vehicle headway distances, the prediction model calculated the likelihood of using one or several lane changes to successfully reach a target position on the road. When approaching a lane drop, the onboard system would use the model to constantly estimate the probability of successfully leaving the blocked lane before reaching the lane-end and advised the driver to start a lane changing maneuver when that probability dropped below a certain threshold. We used the proposed system in a simulation case study on a segment of the I-81 interstate highway with two lane drops - transitioning from four lanes to two lanes - to advise a group of vehicles on avoiding the lane drops. The results showed that the proposed system could effectively reduce average delay, but the reduction depended on the probability threshold, penetration rate of the proposed system, and traffic flow. We concluded that larger probability thresholds are favored for a general case with one lane drop, while traffic simulations are needed to determine the proper probability threshold for more complex cases. We also noted that the proposed system could be simply implemented through in-vehicle navigation systems and could be combined with other methods (such as VSL strategies) for further efficacy. \par
	
	Future work will build upon the results of this study and use a full-cabin driving simulator to analyze the impact of the proposed system on driving behavior. Future research will also examine dynamic assignment of the probability threshold, as well as possible integration of the proposed system with other delay reduction strategies for increased performance.
	
	\section*{Acknowledgment} \label{Section5}
	
	The authors wish to thank the anonymous reviewers for their valuable feedback and constructive criticism of the work. They also wish to express their gratitude to Dr. Harpreet S. Dhillon for his help with the probability model and to Dr. Montasir Abbas and Awad Abdelhalim for their assistance with VISSIM\textsuperscript{\textregistered} simulations.
	
	\bibliography{BIB}
	\bibliographystyle{elsarticle-harv}

	\vfill
	
\end{document}